\documentclass[]{article}
\usepackage[top=2cm, bottom=3cm, left=2.25cm, right=2.25cm]{geometry}
\usepackage{amsmath, subfigure}
\usepackage[small, bf]{caption}
\usepackage[numbers,sort&compress]{natbib}
\usepackage[all=normal, floats]{savetrees}
\usepackage{setspace}
\usepackage{placeins}
\author{Setareh Shahsavari, Gareth H. McKinley\\ \normalsize{Department of Mechanical Engineering, Massachusetts Institute of Technology}}
\usepackage{comment}
\usepackage[final,notref, notcite]{showkeys}
\usepackage{color}
\usepackage{graphicx}
\usepackage{epstopdf}
\usepackage{caption}
\usepackage{sectsty}
\usepackage{placeins}
\usepackage{tikz}
\usepackage[T1]{fontenc}
\usepackage{amssymb}
\usepackage{color,soul}
\sectionfont{\normalsize}
\subsectionfont{\normalsize}
\subsubsectionfont{\normalsize}
\title{\Large Mobility of Power-law and Carreau Fluids through Fibrous Media}
\begin{document}
	\maketitle
	\doublespacing{
	\textbf{\\Abstract}
	The flow of generalized Newtonian fluids with a rate-dependent viscosity through fibrous media is studied with a focus on developing relationships for evaluating the effective fluid mobility. Three different methods have been used here: i) a numerical solution of the Cauchy momentum equation with the Carreau or power-law constitutive equations for pressure-driven flow in a fiber bed consisting of a periodic array of cylindrical fibers, ii) an analytical solution for a unit cell model representing the flow characteristics of a periodic fibrous medium, and iii) a scaling analysis of characteristic bulk parameters such as the effective shear rate, the effective viscosity, geometrical parameters of the system, and the fluid rheology.  Our scaling analysis yields simple expressions for evaluating the transverse mobility functions for each model, which can be used for a wide range of medium porosity and fluid rheological parameters. While the dimensionless mobility is, in general, a function of the Carreau number and the medium porosity, our results show that for porosities less than $\varepsilon\simeq0.65$, the dimensionless mobility becomes independent of the Carreau number and the mobility function exhibits power-law characteristics as a result of high shear rates at the pore scale. We derive a suitable criterion for determining the flow regime and the transition from a constant viscosity Newtonian response to a power-law regime in terms of a new Carreau number rescaled with a dimensionless function which incorporates the medium porosity and the arrangement of fibers.
	\section{Introduction}\label{sec:intro} 

	Many industrial processes involve the flow of non-Newtonian liquids through porous media. Gel permeation chromatography, filtration of polymer solutions and flow of polymer solutions through sand in secondary oil recovery operations are examples of such applications \cite{mullersaezbookchapter1999}. This underlying industrial interest has motivated researchers to develop a variety of tools for analyzing the macro-scale characteristics of the flow, the most important of which is relating the pressure drop per unit length through the porous medium to the volumetric flow rate \cite{bear2013dynamics}. \\
	Predicting the pressure drop\textendash velocity relationship from detailed analysis of the fluid flow at the pore level is computationally expensive but possible for Newtonian fluids \cite{torquato2013random}. The non-Newtonian characteristics of the fluid being pumped add another level of complexity to the system and restrict detailed solution of the equations of motion. Therefore, upscaling and homogenization approaches are commonly used to develop correlations for the hydrodynamic resistance of porous media, in which all the microstructural details of the porous matrix are absorbed into bulk parameters that reflect the average properties of the medium \cite{hornung2012homogenization}.\\
	For the case of a Newtonian fluid, Darcy's law is the constitutive relationship that relates the pressure drop across a porous bed to the apparent fluid velocity \cite{bear2013dynamics}. Darcy's law can be extended to the case of a generalized Newtonian fluid by using the effective viscosity of the fluid \cite{chhabra1993bubbles}. For the flow of power-law fluids in a homogeneous porous medium, an effective or characteristic shear rate, $\dot{\gamma}_{\text{eff}}$, can be defined which leads to an effective value of the rate-dependent viscosity. Therefore it is possible to modify Darcy's law by replacing the constant Newtonian viscosity with an effective viscosity, $\eta_{\text{eff}}$, that depends on the flow rate, porosity (typically denoted by $\varepsilon$), and other system parameters. For a generalized Newtonian fluid, the modified Darcy equation can be written as
			\begin{align}
				\frac{\Delta p}{L}=\frac{\eta_{\text{eff}} U}{\kappa}
				\label{eqn:Darcy-law}
			\end{align}
	where $\Delta p/L$ is the pressure drop per unit length of the porous medium, $U$ is the superficial or apparent velocity of the fluid, and $\kappa$ is the permeability of the porous medium. 
	Since both $\kappa$ and $\eta_{\text{eff}}$ are functions of the porosity of the medium, it is pedagogically convenient to combine the two parameters and introduce the concept of an \textit{effective mobility} which is defined as the ratio of the permeability to the effective viscosity and has units of $m^2Pa^{-1}s^{-1}$ \cite{dullien2012porous}:
			\begin{align}
				M&\equiv \frac{\kappa(\varepsilon)}{\eta_{\rm{eff}}\left(\dot{\gamma}_{\text{eff}},\varepsilon\right)} \label{eqn:mobility-dfn}
			\end{align}
	It is not possible to give a general form of Darcy's law for all complex fluids. White \cite{White1967} discusses the conditions for expecting a similarity solution for non-Newtonian flow through porous media and concludes that for viscoelastic fluids, one cannot find a general solution for calculating pressure drop as a function of flow rate from rheological parameters. Therefore, the majority of studies on flow of viscoelastic fluids through porous media are primarily empirical in nature \cite{Marshall1967,Galindo2012}. \\
	A review of the flow of non-Newtonian fluids in fixed and fluidised beds is given by Chhabra et al. \cite{Chhabra2001}, with a focus on the prediction of macro-scale flow phenomena such as the pressure drop\textendash flow rate relationship. In particular, they review four categories of model: the capillary bundle approach, models based on particulate drag theories, methods based on the volume averaging of the governing field equations, and purely empirical models. A more recent review paper on the flow of non-Newtonian fluids in porous media is presented by Sochi \cite{Sochi2010}, in which the main approaches for describing the flow are categorized as homogenization models, \cite{bourgeat1996,blunt2001,Pearson2002,auriault2002}, numerical methods involving a detailed description of the porous medium at pore-scale \cite{Aharonov1993,porter2005,schulz2007}, and pore-network modeling, which partly captures the characteristics of the porous structure at pore level with affordable computational resources \cite{Sochi2010,blunt2001, sinha2007}. A recent work using drag theories to calculate the effective viscosity of complex fluids is by presented by Housiadas and Tanner \cite{housiadas2014} for suspensions of spherical particles, where the viscous force exerted by a fluid flowing on a particle is calculated in a Brinkman medium as a function of the volume fraction.\\
	For the flow of power-law fluids, the capillary bundle model is the most common approach used to predict the pressure drop resulting from flow through porous media \cite{Christopher1965,Mishra1975,Wang1979}. In this method, the pore structure is modeled as a bundle of capillaries which gives rise to the same hydrodynamic resistance as the porous medium. The flow is considered to be fully developed in the capillaries, and an empirical tortuosity factor is introduced to take into account the microstructural complexities of the porous medium. A clear description of this modeling technique is presented by Bird et al. \cite{Bird-transport}. When the fluid flowing through the pore space is Newtonian, this approach leads to the well-known Blake-Carman-Kozeny equation, which is the most commonly used equation for packed beds \cite{ergun1952}. Christopher and Middleman \cite{Christopher1965} used a capillary model to develop a modified Blake-Carman-Kozeny equation for the laminar flow of inelastic power-law fluids through granular porous media. They also experimentally investigated their model for the flow of dilute solutions of carboxymethylcellulose through a tube closely-packed with small glass spheres. Using a similar approach, Mishra et al. \cite{Mishra1975} related the average shear stress to the average shear rate to predict the flow behaviour of power-law fluids and used polyvinyl alcohol solutions in water as a representative of power-law fluids for experimental verification of their model. Duda et al. \cite{Duda1983} discuss the limitations of the capillary model based on experimental studies of the flow of inelastic solutions in porous media and suggest that the rheological model for the fluid must include the characteristic transition from Newtonian behavior at low shear rates to shear-thinning behavior at high shear rates. Other modeling techniques that have been used less commonly for power-law fluids in the literature include volume averaging of the equations of motion \cite{Smit1999} and pore network simulations \cite{Shah1995}. \\
	The majority of these models are developed for flow through fixed beds of spherical particles. There has been less focus on the flow of non-Newtonian fluids through fibrous media. The anisotropic characteristics of fibrous media can lead to a markedly different fluid behavior compared to packed beds of particles \cite{Bruschke1993}. As a result, relationships that are developed for isotropic media are inadequate to predict the flow behavior in fibrous media. Bruschke and Advani \cite{Bruschke1993} derived an analytical relationship for the mobility of power-law fluids through fibrous media using the lubrication theory to describe the flow transverse to an array of cylinders. Their relationship can predict the mobility in the limit of low porosity, where the cylinders are relatively close and the assumptions for the lubrication theory hold. A more general analysis is required to predict the mobility for wider ranges of porosity.\\
	In this study we develop analytical and scaling models for quantifying the mobility of power-law fluids through fibrous media. In the following sections, we first study the problem numerically using pore-scale solution of the equations of motion in an idealized geometry (consisting of a periodic array of cylinders) to represent the fibrous medium and investigate the effects of different system parameters on the mobility of the fluid. Then, we develop analytical expressions by solving the equation of motion for the creeping flow of power-law fluids transverse to a confined cylinder with appropriate boundary conditions to capture the periodicity of the system. By employing a scaling analysis of the flow through the porous network, we propose a theoretical model that can be used as a modified Darcy law. In addition to the simple Ostwald-de Waele power-law fluid, we also investigate Carreau fluids \cite{Bird1987} using our numerical simulations and explore the parameters that affect the transition from Newtonian to power-law behavior. Finally, we extend our scaling model that was developed for power-law fluids to predict the pressure drop\textendash velocity relationship for flow of Carreau fluids through fibrous media.
	\section{Governing equations}\label{sec:governing eqns}
	Power-law fluids are a sub-category of generalized Newtonian fluids, for which the constitutive response is inelastic and the viscosity,  $\eta$, is directly proportional to a power of the characteristic deformation rate $\dot{\gamma}$ with the power-law exponent depending on the material composition and concentration \cite{Bird1987}.  The following equations, respectively, represent the constitutive law for generalized Newtonian fluids and the definition of the characteristic shear rate in terms of the second invariant of the velocity gradient tensor. 
		\begin{align}
			\boldsymbol{\tau}&=-\eta\left(\dot{\gamma}\right) \boldsymbol{\dot{\gamma}}=-\eta\left(\dot{\gamma}\right)\left\{  \nabla\textbf{u}+\nabla\textbf{u}^T\right\}
			\label{eqn:GNF-stress}\\
			\dot{\gamma}&=\sqrt{\frac{1}{2}\,\left(\boldsymbol{\dot{\gamma}}:\boldsymbol{\dot{\gamma}}\right)}
			\label{eqn:shear rate definition}
		\end{align}	
	Here, $\boldsymbol{\tau}$ is the stress tensor, $\textbf{u}$ is the velocity vector, and $\boldsymbol{\dot{\gamma}}=\nabla\textbf{u}+\nabla\textbf{u}^T$ denotes the deformation rate tensor and we use the sign convention of \cite{Bird1987}.  The power-law model of Ostwald and de Waele for the viscosity function is
		\begin{align}
			\eta&\left(\dot{\gamma}\right)=m\dot{\gamma}^{n-1}\label{eqn:PL-viscosity}
		\end{align}
	where $n$ and $m$ denote the power-law exponent and the consistency of the fluid respectively. A power-law exponent of $0<n<1$ represents a shear-thinning fluid, while $n>1$ shows that the fluid is shear-thickening. The simple power-law model (equation(\ref{eqn:PL-viscosity})) has a well-known singularity at zero shear rate, which must be carefully accounted for in kinematic analyses. The Carreau-Yasuda equation is an alternate generalized Newtonian model that enables the description of the plateaus in viscosity that are expected when the shear rate is very small or very large \cite{Bird1987}. The shear viscosity function for this model is given by
		\begin{align}
			\frac{\eta\left(\dot{\gamma}\right)-\eta_\infty}{\eta_0-\eta_\infty}=\left(1+\left(\lambda\dot{\gamma}\right)^2\right)^{\frac{n-1}{2}},
			\label{eqn:Carreau-viscosity}
		\end{align}	 
	where $\eta_0$ is the zero-shear-rate viscosity, $\eta_\infty$ is the infinite-shear-rate viscosity, $\lambda$ is the inverse of a characteristic shear rate at which shear thinning becomes important, and $n$ is the power-law exponent as before. Many complex fluids used in enhanced oil recovery (e.g. worm-like micellar solutions) are well described by this model \cite{Haward2012}. \\
	For the flow of a power-law fluid through a homogeneous porous medium, an effective or characteristic shear rate can be defined, which depends on the porosity, pore size, apparent velocity, and power-law exponent, as derived by Christopher and Middleman \cite{Christopher1965} for the case of packed spheres. The effective viscosity for power-law or Carreau fluids can then be obtained by substituting this effective shear rate into equations (\ref{eqn:PL-viscosity}) or (\ref{eqn:Carreau-viscosity}), respectively. This effective viscosity determines the mobility of the fluid flowing through a porous medium of known permeability based on equation (\ref{eqn:mobility-dfn}).\\
	The Blake-Carman-Kozeny equation is the most common semi-empirical expression that is used to describe the laminar flow of Newtonian fluids through granular media and can be written in the form:
		\begin{align}
			\kappa=\frac{D_p^2\varepsilon^3}{150 \left(1-\varepsilon\right)^2},
			\label{eqn:permeability-BCK}
		\end{align}
	where $D_p$ is the effective grain size, $\varepsilon$ is the porosity, and the numerical factor of 150 was originally determined by comparison with experimental data. Christopher and Middleman \cite{Christopher1965} used a modified Blake-Carman-Kozeny equation for the laminar flow of power-law fluids through a packed bed, which was based on the equation given by Bird et al. \cite{Bird-transport} but with a different tortuosity coefficient. In addition, Christopher and Middleman suggested that the wall shear rate can be estimated by the following equation, which is derived based on the capillary model by considering steady flow of a power-law fluid in a long cylindrical tube representing the hydrodynamic resistance of the porous medium:
		\begin{align}
			\dot{\gamma}_w=\frac{3\left(3n+1\right)}{n}	\frac{U}{\sqrt{150\,\kappa\, 
			\varepsilon}} \label{eqn:shear_middleman}				
		\end{align}	
	Like the majority of capillary bundle models, this semi-empirical equation is developed for applications in granular media. A model more specific to fibrous media was developed by Bruschke and Advani \cite{Bruschke1993}. They considered flow of power-law fluids transverse to a periodic array of cylinders as schematically shown in Figure \ref{fig:cylinder-array}. By invoking symmetry arguments they solved the equations of motion in the unit cell shown in Figure \ref{fig:unit_cell} and applied the lubrication approximation to obtain a closed form analytical solution that describes the transverse flow through fibrous media at low porosities. For a periodic array of cylinders having diameter $d$ and center-to-center spacing $s$, the mobility from the lubrication solution derived by Bruschke and Advani \cite{Bruschke1993} is given as
			\begin{align}
				M&=\frac{1}{4} \frac{U^{1-n} s^{1+n}}{m}\left(\frac{n}{1+2n}\right)^n\frac{1}{\sqrt{\pi}}\left[\left(1+\frac{d}{s}\right)^{-(2n+1)}\left(\cos^2{\alpha}\right)^{2n+1}\alpha^{-4n-1} \frac{\Gamma\left(2n+1/2\right)}{\Gamma\left(2n+1\right)}\right]^{-1}
				\label{eqn:mobility-Bruschke}
			\end{align}	
		where $\Gamma$ denotes the gamma function and the geometric parameter $\alpha$ is only a function of the relative fiber spacing (or porosity):
			\begin{align}
				\alpha=\arctan\left(\frac{1-d/s}{1+d/s}\right)^\frac{1}{2}.
			\end{align}
	This model predicts the mobility of power-law fluids in fibrous media with a square arrangement of fibers in the limit of low porosity where the lubrication assumptions are valid. We will later show a this model with our numerical results. Here, we develop a more comprehensive mobility function that covers a wider range of porosities and fiber arrangements by employing insight from numerical simulations of the same system.\\
	Based on the dimensionality of the mobility, we can define a dimensionless mobility function for power-law fluids in the form:
		\begin{align}
			M^*&\equiv\frac{\kappa}{\eta_{\text{eff}}}\left(\frac{mU^{n-1}}{d^{n+1}}\right) \label{eqn:mobility-nd-dfn}
		\end{align} 
	The corresponding form for the Carreau model is
		\begin{align}
			M^*&\equiv\frac{\kappa}{\eta_{\text{eff}}} \left(\frac{\eta_0-\eta_\infty}{d^2}\right)Cu^{n-1} \label{eqn:mobility-nd-dfn-Carreau}
		\end{align} 
	where $Cu$ is the Carreau number defined as
		\begin{align}
			Cu=\frac{\lambda U}{d}.
		\end{align}
	Unlike the dimensional form of the mobility, the dimensionless mobility does not explicitly depend on the flow rate; it is a function of porosity, $\varepsilon$, power-law exponent, $n$, and the rheology of the fluid. Therefore, we will present our results for the mobility function through a fiber bed using the dimensionless forms defined above corresponding to power-law or Carreau fluids.\\ 
	The equations governing the motion of an incompressible generalized Newtonian fluid are the Navier-Stokes equations with a shear-rate-dependent viscosity function, $\eta\left(\dot{\gamma}\right)$ that is given by equation (\ref{eqn:PL-viscosity}) or (\ref{eqn:Carreau-viscosity}) or other inelastic constitutive relationship. The continuity and momentum equations for the steady state incompressible flow are as follows:
		\begin{align}
			\nabla.\,\textbf{u}               & =0 \label{eqn:Continutity}                                                                                                               \\
			\rho\textbf{u}\,.\nabla\textbf{u} & =-\nabla p + \nabla.\left(\eta\left(\dot{\gamma}\right) \left\{  \nabla\textbf{u}+\nabla\textbf{u}^T\right\}\right) \label{eqn:Momentum}
		\end{align}
		\begin{figure}[hbtp]
		  \centering
		  	\subfigure[]{
	                \includegraphics[width=.4\textwidth]{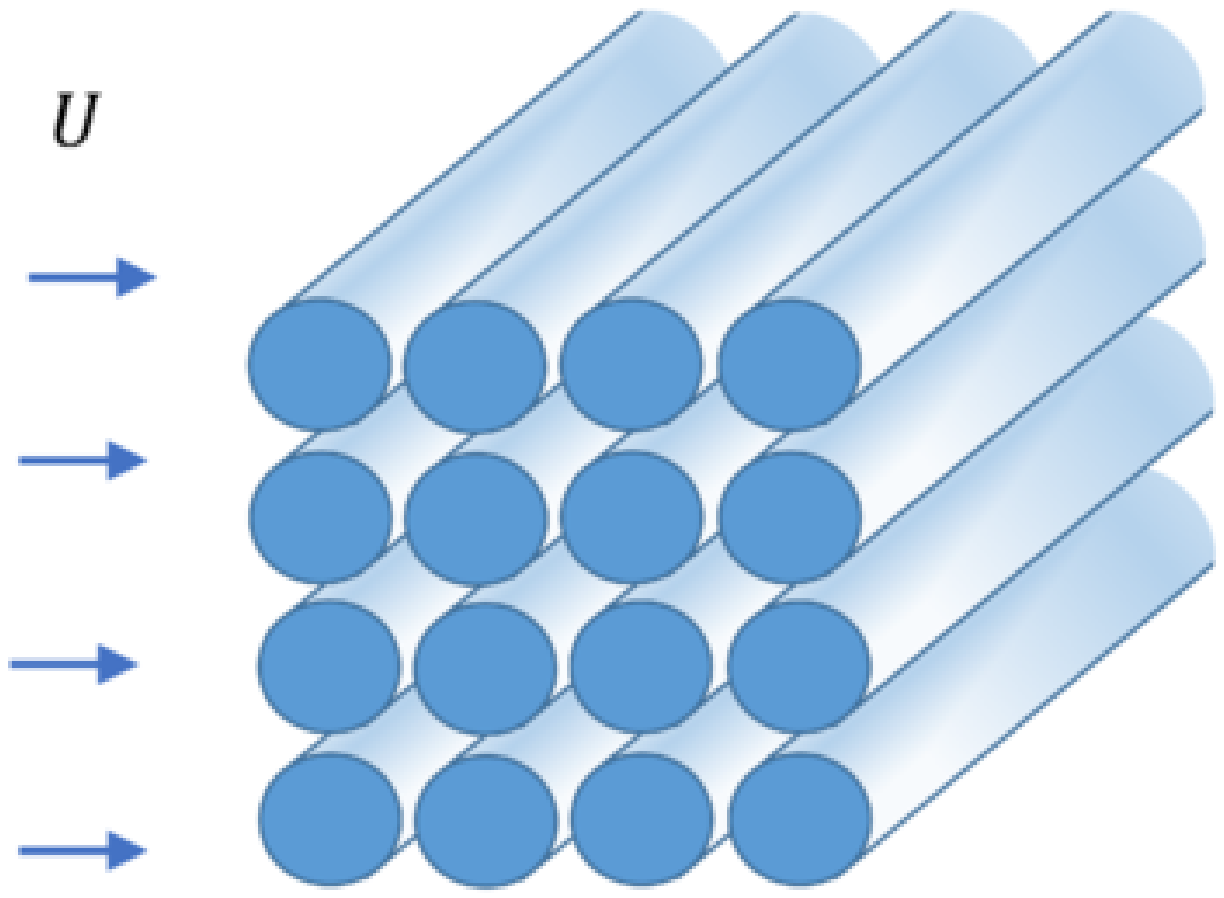}
	                \label{fig:cylinder-array}
	        }
	        \subfigure[]{
	                \includegraphics[width=.4\textwidth]{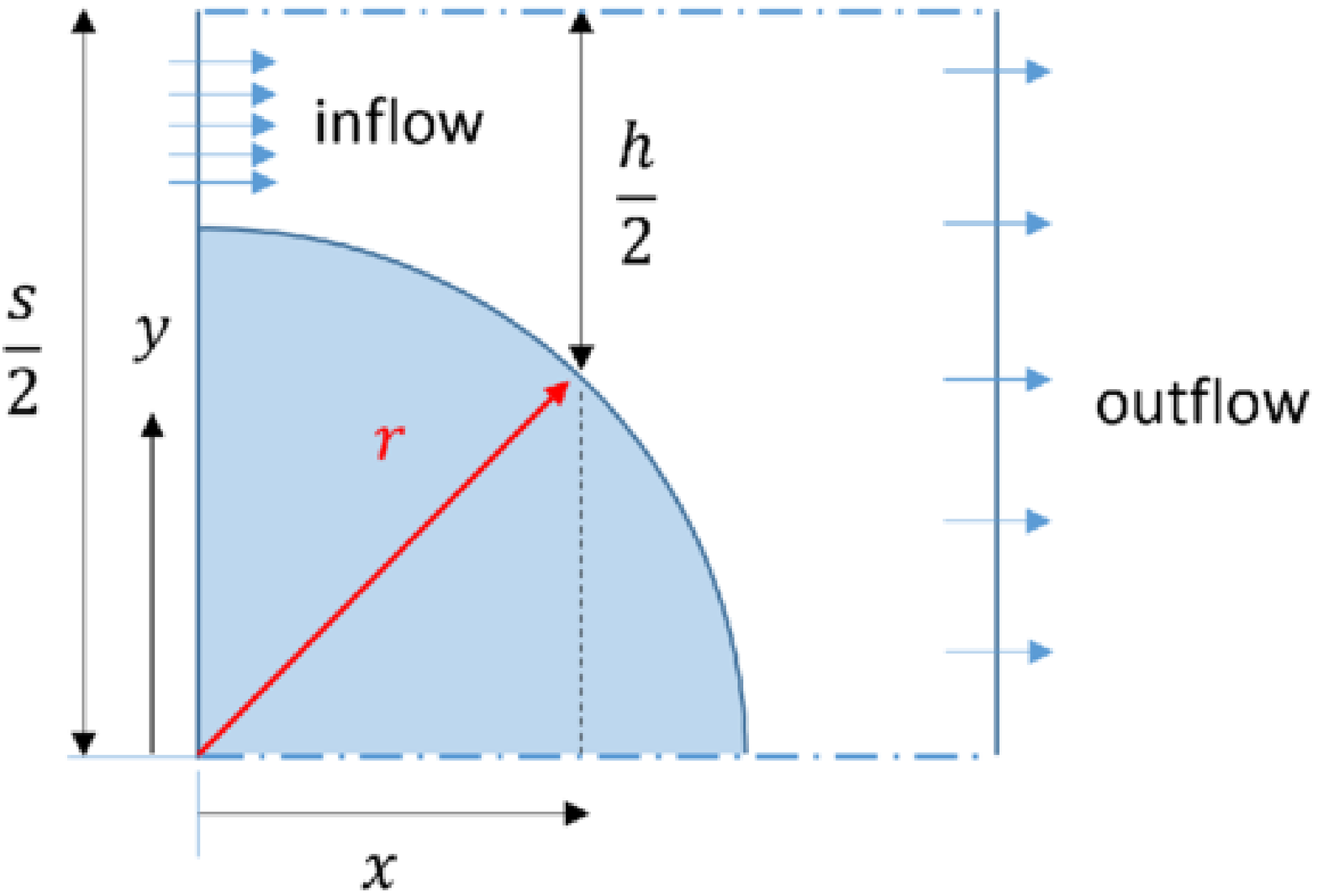}
	                \label{fig:unit_cell}
	                }
		  \caption[fig:flow_cylinders]
		   {(a) Schematic of flow transverse to a periodic array of cylinders (b) unit cell used for the lubrication model developed by Bruschke and Advani \cite{Bruschke1993}.}
		\end{figure}		 
		\begin{figure}[h!]
		  \centering
		  \subfigure[]{
		  	\includegraphics[width=0.35\textwidth]{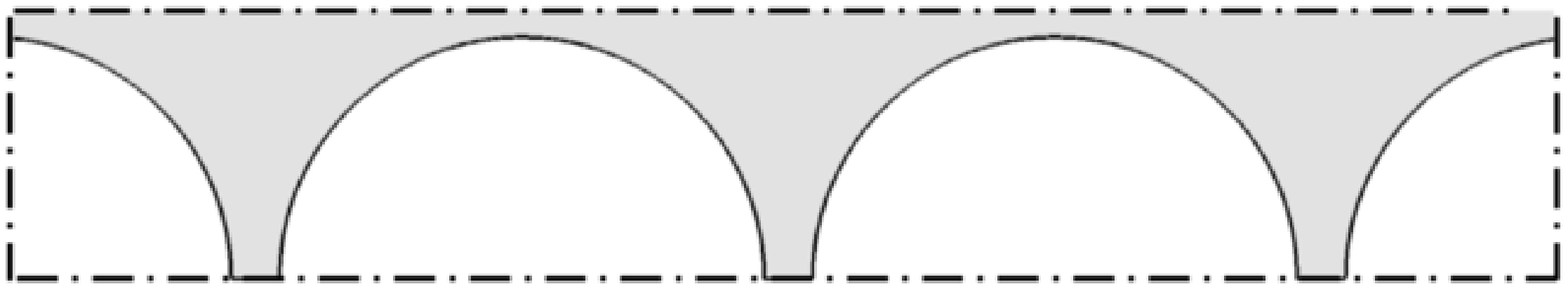}
		  	\label{fig:domain-square}
		  }\\
		  \subfigure[]{
		  	\includegraphics[width=0.35\textwidth]{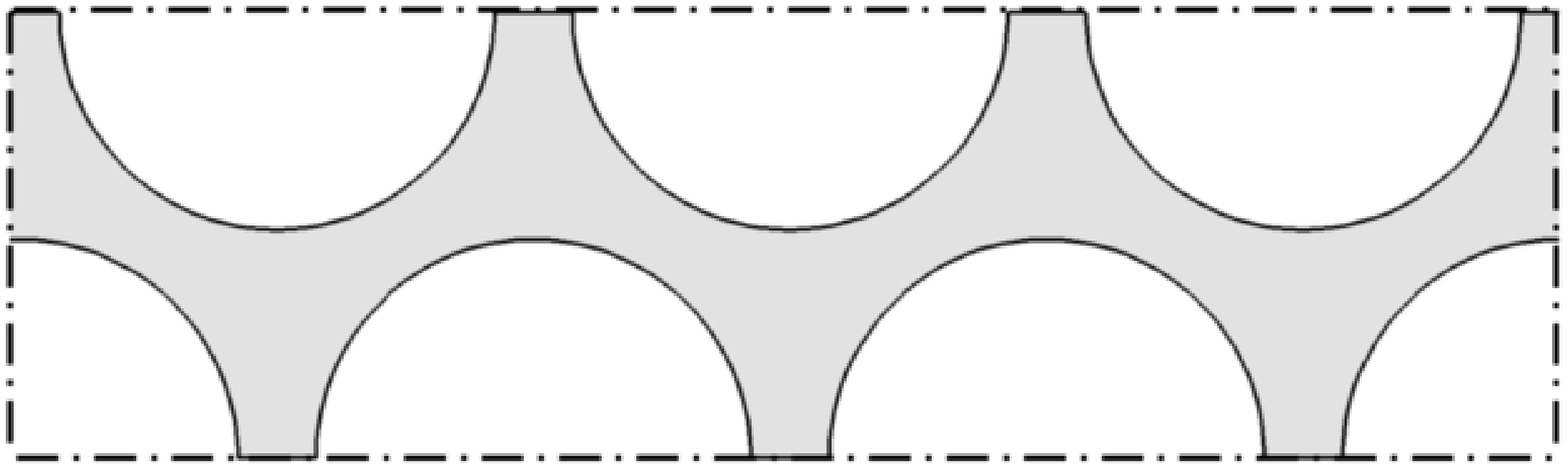}\label{fig:domain-hex}
		  	}\\
		  \label{fig:domain}
		  \caption[]
		   {Schematic of the flow domains studied numerically: (a) geometry for a square arrangement of cylinders (b) hexagonal arrangement . Both geometries represent a porosity of $\varepsilon=0.35$. Boundary conditions assumed on the left and right faces are periodic and on the top and bottom faces are symmetric.}	 
		   \label{fig:domain}  
		\end{figure}
\FloatBarrier		
	In order to model the flow of power-law fluids through fibrous media, we consider an idealized domain consisting of a periodic array of cylinders as shown schematically in Figure \ref{fig:cylinder-array} with two different lattice arrangements: square and hexagonal. These two geometric arrangements (schematically shown in Figure \ref{fig:domain-square} and \ref{fig:domain-hex}) are selected since they offer the maximum and minimum tortuosity for a homogeneous medium consisting of parallel cylinders of constant diameter. The set of equations (\ref{eqn:PL-viscosity}), (\ref{eqn:Continutity}), (\ref{eqn:Momentum}) are solved numerically using COMSOL Multiphysics 4.3a for a wide range of relative fiber spacings and rheological parameters. Additional details of the numerical model and results are given in section \ref{sec:numerical}. In this section, we present the results for the mobility of power-law fluids in fibrous media to show how it compares with the existing theoretical or empirical relations.\\
	Figure \ref{fig:mobility-capil-lubr} compares the results of the modified capillary model \cite{Christopher1965} and the lubrication model \cite{Kuwabara1959} with our numerical calculations for square and hexagonal arrangement of fibers and a fluid power-law exponent of $n=0.5$. In this figure, for the capillary model (dashed line), we used the Blake-Carman-Kozeny permeability equation (\ref{eqn:permeability-BCK}) along with the effective viscosity obtained from the wall shear rate, equation (\ref{eqn:shear_middleman}), derived by Christopher and Middleman \cite{Christopher1965}.  Note that the minimum porosity for an array of cylindrical fibers is $\varepsilon_{\text{min}}=1-\pi/4\simeq0.21$ for square arrangement and $\varepsilon_{\text{min}}=1-\pi/(2\sqrt{3})\simeq0.09$  for hexagonal arrangement of fibers. It is clear from the numerical simulations shown in Figure (\ref{fig:mobility-capil-lubr}) that the specifics of fiber arrangement (i.e. square or hexagonal) becomes more significant at low porosities ($\varepsilon<0.4$). In this limit, the lubrication model describes the square arrangement with reasonable accuracy (due to the additional symmetry of the flow in this configuration) while the modified capillary model better predicts the hexagonal arrangement. The errors from the capillary model and from the lubrication model at $\varepsilon=0.5$ are 17\% and 30\% respectively. For the lubrication model, the error grows as the porosity increases while for the capillary model, the error does not change monotonically and is minimum for intermediate porosities ($0.5<\varepsilon<0.8$) since it involves an empirical tortuosity factor.
		\begin{figure}[]
			  \centering
			  \includegraphics[width=0.5\textwidth]{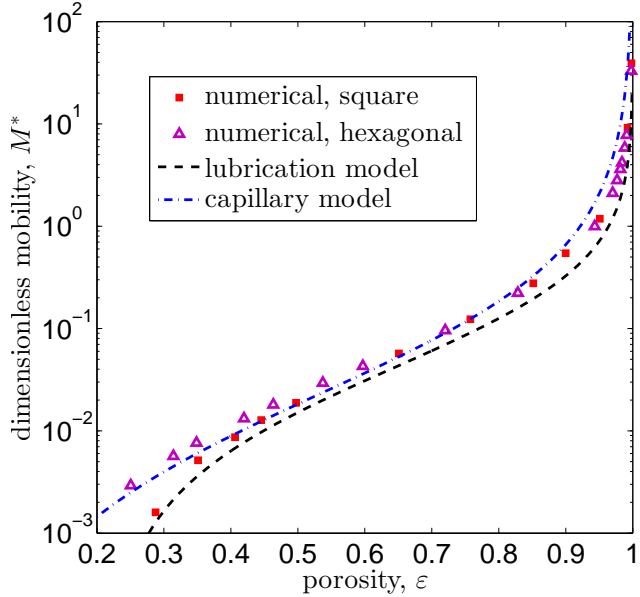}
			  \caption[]
			  {Comparison of mobility models with numerical solution for flow of a power-law fluid with power-law exponent of $n=0.5$ through fibrous media with square and hexagonal arrangement of fibers. The lubrication model is derived by Bruschke \& Advani (1993) and the modified capillary model is based on the effective shear rate derived by Christopher \& Middleman (1965).}
			  \label{fig:mobility-capil-lubr}
		\end{figure}
\FloatBarrier
	Comparison of the computational data in Figure \ref{fig:mobility-capil-lubr} shows that although the existing analytical and semi-empirical solutions describe the general trend observed in the data very well, they are not sufficiently accurate to predict the mobility over the entire range of porosity. The objective of the present study is to develop theoretical solutions to establish a modified Darcy equation for the flow of power-law fluids through fibrous media that can be used to accurately calculate the mobility for a wide range of porosity and rheological parameters (i.e. $m$ and $n$). To achieve this, we take two theoretical approaches: analytical solution of the equation of motion in the geometry shown in Figure \ref{fig:unit_cell} and scaling analysis. To derive an analytical solution, we consider the Cauchy momentum equation in cylindrical coordinates and assume locally fully-developed laminar flow. For the scaling solution, we derive an effective viscosity in the fibrous medium as a function of the system parameters based on appropriate scaling estimates for the characteristic magnitude of the shear rate $\dot{\gamma}_{\text{eff}}$ in the unit cell as the geometric arrangement and the cylinder spacing $s$ change. We verify our analytical and scaling results with full numerical computations of the described inelastic model. The details of our numerical simulations are presented in section \ref{sec:numerical} below.
\FloatBarrier
	\section{Numerical approach}\label{sec:numerical}
	We solve the system of equations introduced in section \ref{sec:governing eqns} (equations (\ref{eqn:Continutity}) and (\ref{eqn:Momentum})) numerically for the two-dimensional flow of power-law fluids and Carreau fluids past an array of circular cylinders arranged in square and hexagonal patterns. COMSOL Multiphysics 4.3a is used to build the numerical model and solve the equations. In Figure \ref{fig:domain-square} and (b), sections of the flow domain are shown for the square and hexagonal arrangement respectively, both of which have a porosity of $\varepsilon=0.35$. The complete domain contains 16 repeated unit cells (each cell having a width of one cylinder spacing, $s$) and the length of the inlet and outlet regions are long enough (each 16 times the cylinder spacing) to eliminate the entrance and exit effects. Depending on the relative fiber spacing, different number of mesh elements ($8\times10^4-1.2\times10^5$) are used to obtain a mesh-independent result. The following boundary conditions are used: symmetry at the top and bottom surface, uniform velocity at the inlet, and zero relative pressure at the outlet. \\
	We have conducted parametric studies to investigate the effects of the flow rate ($Q$), fiber spacing ($s$) and diameter ($d$), power-law exponent ($n$), and Carreau number ($Cu$) independently. From the numerical values of the pressure drop across the domain, we compute the permeability and mobility, the results of which will be shown and discussed in the subsequent sections. In this section we first discuss local dynamical features (e.g. stress profiles, contours, and streamlines) that cannot be captured in detail by the global parameters that appear in theoretical models.\\
	Figure \ref{fig:stress} shows the distribution of the tangential viscous stress on a single fiber for two different porosities, $\varepsilon=0.35$ and $0.85$. All other parameters are kept constant and only the fiber spacing is varied. In figures \ref{fig:stress_N} and \ref{fig:stress_PL}, we have used the apparent velocity, viscosity, and fiber diameter to non-dimensionalize the viscous stress (the characteristic viscous stress is $\eta U/d$ for Newtonian fluids and $mU^n/d^n$ for power-law fluids). As the figure shows, there are several orders of magnitude difference between the scale of the stress in the high porosity and low porosity cases for both the Newtonian and power-law fluid. To capture the primary effects of porosity, in figures \ref{fig:stress_N_scaled}  and \ref{fig:stress_PL_scaled}, we rescale the stresses using a dimensionless coefficient (derived in section \ref{sec:scaling analysis} and given by equation (\ref{eqn:gamma*})) such that the plots are now more appropriately scaled and have the same order of magnitude for different porosities and fluids. These plots show the importance of geometric factors on the scaling of the stress, and we discuss the scaling analysis further in section \ref{sec:scaling analysis}.
	\\ Also in Figure \ref{fig:stress}, we can see there are regions near $\theta\approx0, \pi$ where the shear stress on the wall changes sign. This is due to formation of weak recirculating vortices in the gap between the fibers at low porosities. In Figure \ref{fig:streamlines-square}, we have plotted the changes in the streamlines that arise for different porosities to show the effect of varying the fiber spacing on the appearance of the vortices. The strength of these vortices are, however, very low and when streamlines are drawn based on fixed increments in stream function magnitude (e.g. the top figure in Figure \ref{fig:streamlines-square}), they are not observable and the fluid between the cylinders appears to be quiescent. In the bottom figures the streamlines are drawn based on a uniform density to better reveal the structure of the weak recirculating eddies.\\ 
	In Figures \ref{fig:velocity-viscosity-hex} (a) and (b) respectively, we present the contours of the velocity magnitude and the logarithm of the dimensionless local viscosity distribution, $\log\left(\eta\, m^{-1}\left(U/d\right)^{1-n}\right)$ for transverse flow over cylinders in a hexagonal arrangement with porosity $\varepsilon=0.18$. Based on the numerical simulation, at this small value of porosity the local viscosity varies over three orders of magnitude within the porous medium. In Figure \ref{fig:velocity-viscosity-hex} (c), we show the numerical results for the velocity magnitude in the gap between two cylinders (along the white vertical line shown in (a)). This plot shows that in the recirculating regions, the velocity is several orders of magnitude less than the bulk velocity $U$ and alternates in direction in each recirculating vortex.\\
	\begin{figure}[h!]
	        \centering
	        \subfigure[]{
	                \includegraphics[width=.4\textwidth]{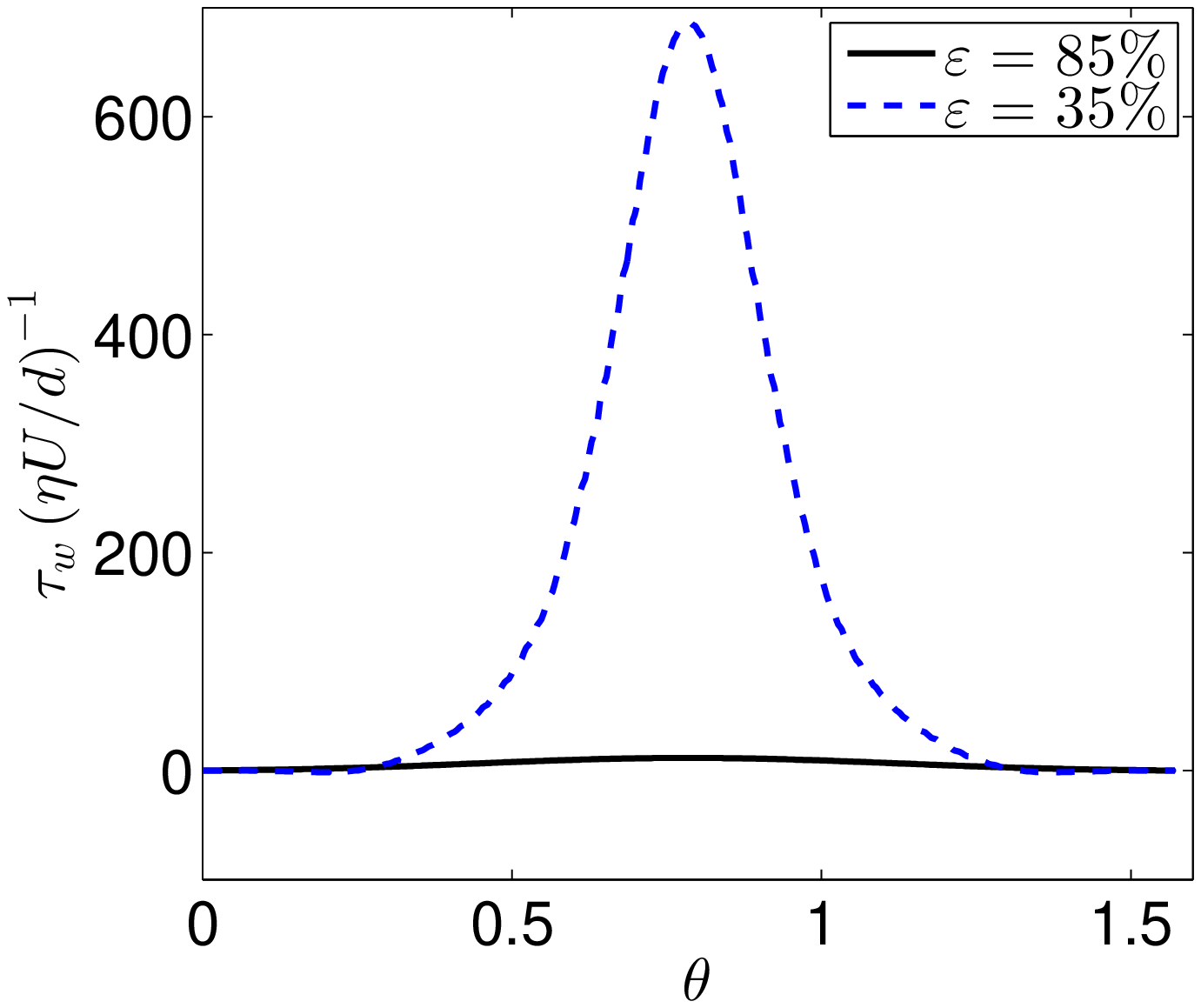}
	                \label{fig:stress_N}
	        }
	        \subfigure[]{
	                \includegraphics[width=.4\textwidth]{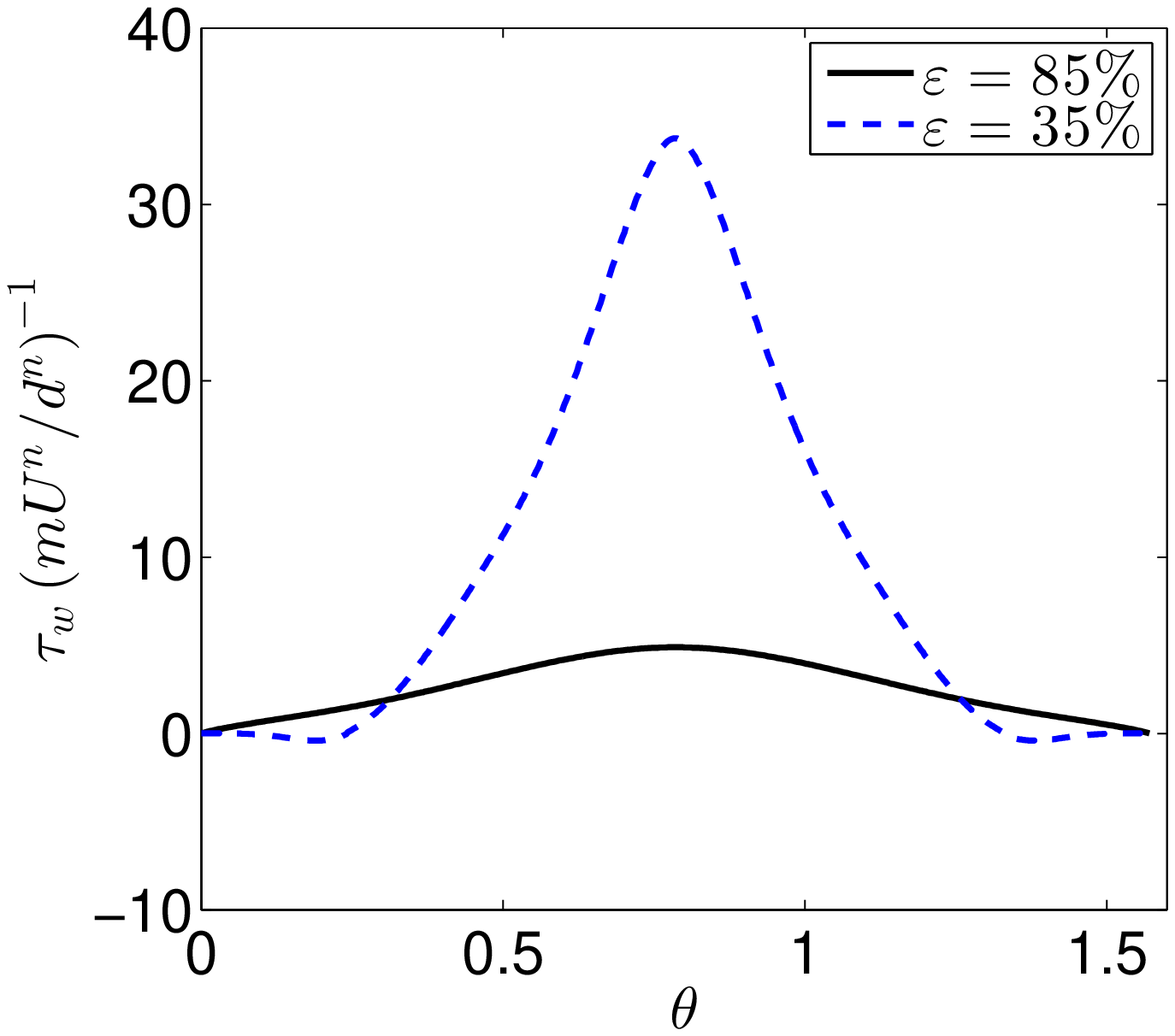}
	                \label{fig:stress_PL}
	                }
	        \subfigure[]{
	                \includegraphics[width=.4\textwidth]{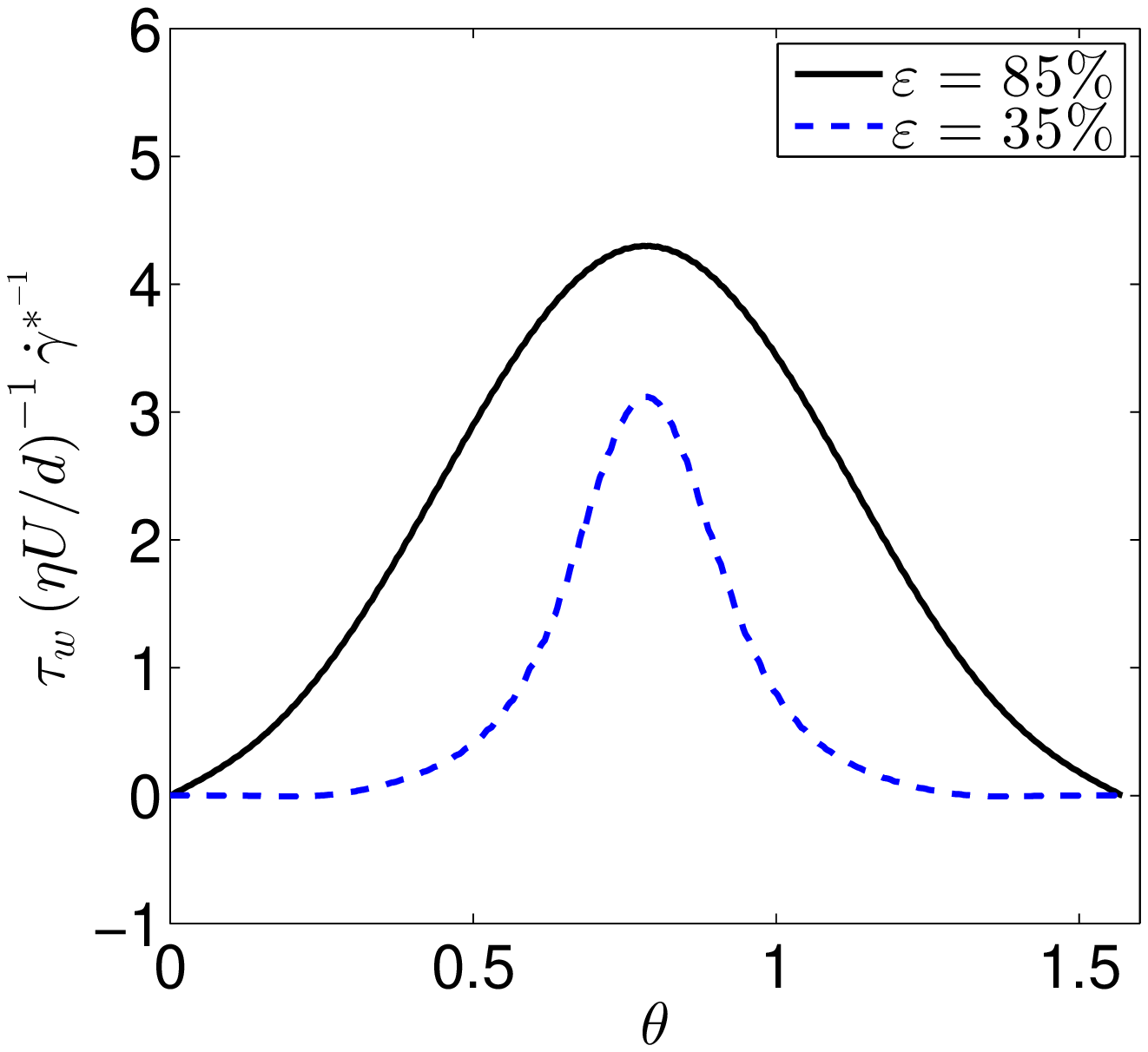}
	                \label{fig:stress_N_scaled}
	        }
	        \subfigure[]{
	                \includegraphics[width=.4\textwidth]{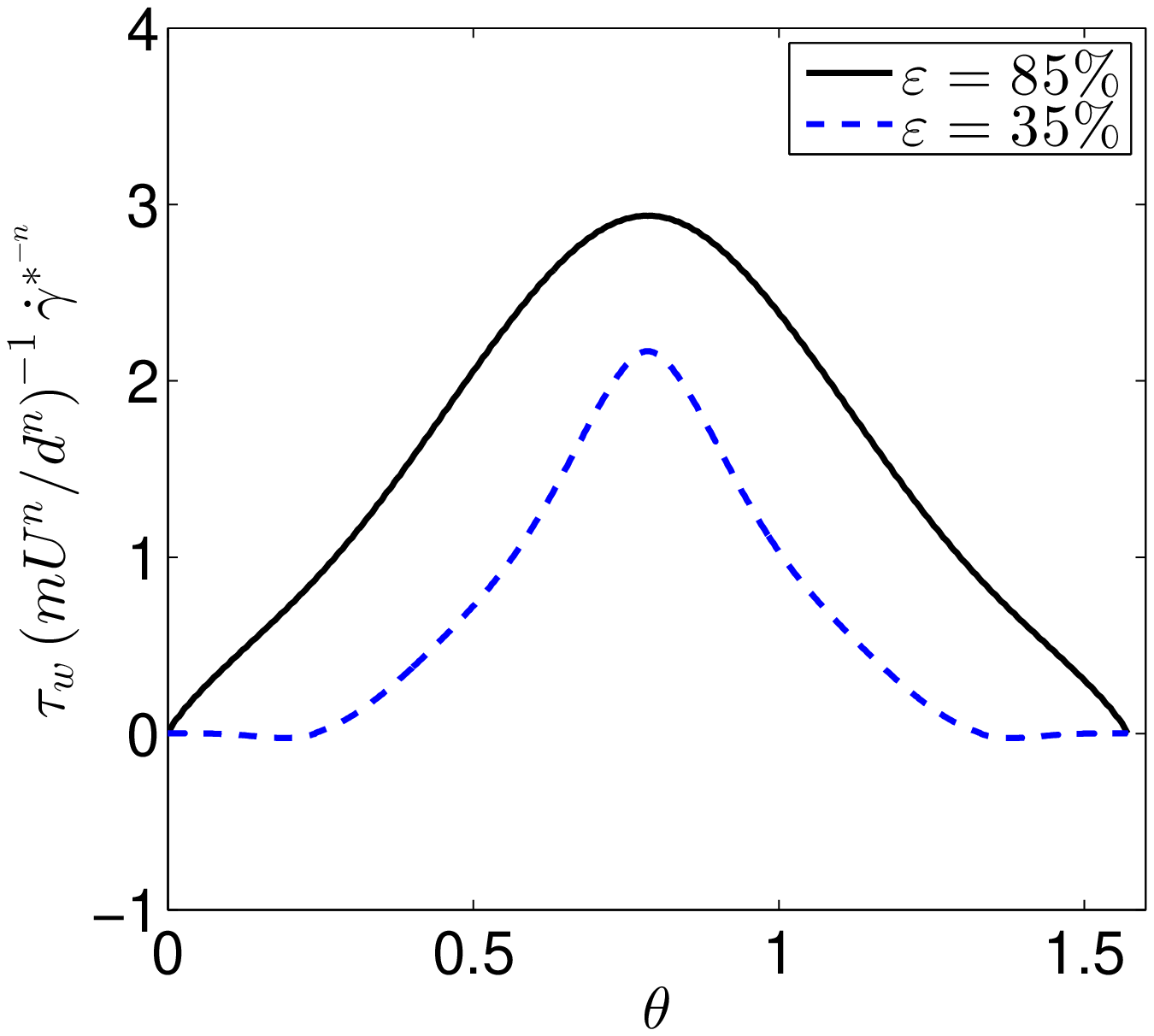}
	                \label{fig:stress_PL_scaled}
	        }
	        \caption{Distribution of the dimensionless wall shear stress around a fiber: (a) dimensionless shear stress for a Newtonian fluid $\tau_{N}=\tau_w/\left(\eta U/d\right)$, (b) for a power-law fluid $\tau_{p}=\tau_w/\left(mU^n/d^n\right)$, (c) rescaled dimensionless shear stress for a Newtonian fluid $\tau^*_N=\tau_{N}/\dot{\gamma}^*$, (d) for a power-law fluid $\tau^*_p=\tau_{p}/\dot{\gamma}^{*^n}$, using the coefficient $\dot{\gamma}^*$ given by equation (\ref{eqn:gamma*}).}
	        \label{fig:stress}
	\end{figure} 
		\begin{figure}[h!]
			  \centering
			  \includegraphics[width=0.7\textwidth]{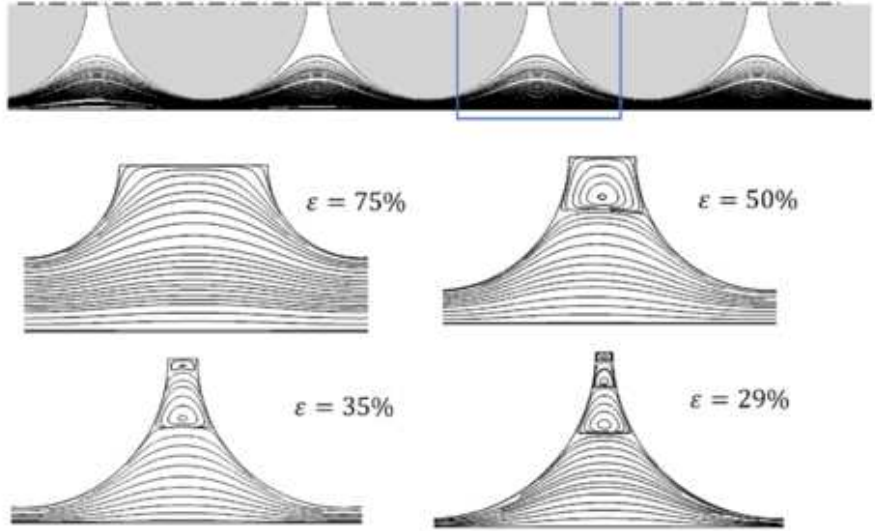}
			  \caption[]
			  {Streamlines for the flow of a shear-thinning fluid with power-law exponent $n=0.5$ for different fiber spacings. The top figure shows 100 streamlines with equal spacing between the minimum and maximum values. In the four bottom figures (zoomed into the region shown by the blue box in the upper image) the streamlines are drawn with uniform density throughout the domain.}
			  \label{fig:streamlines-square}
		\end{figure}
		 \begin{figure}[!htb]
		    \centering
		    \begin{minipage}[b]{0.45\textwidth}
		      \begin{subfigure}[]{
    		    \includegraphics[width=0.9\textwidth]{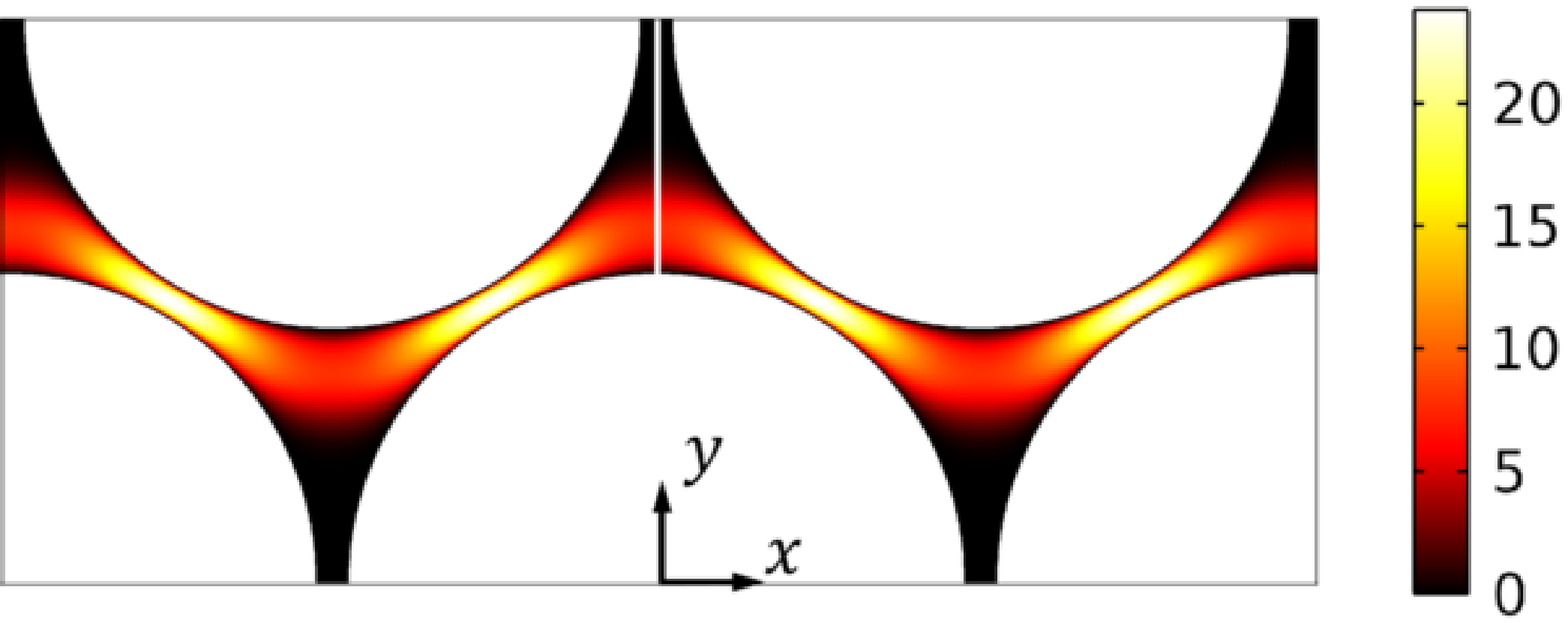}
    		    		    }
		      \end{subfigure}\\[\baselineskip]
		      \begin{subfigure}[]{
    		    \includegraphics[width=0.9\textwidth]{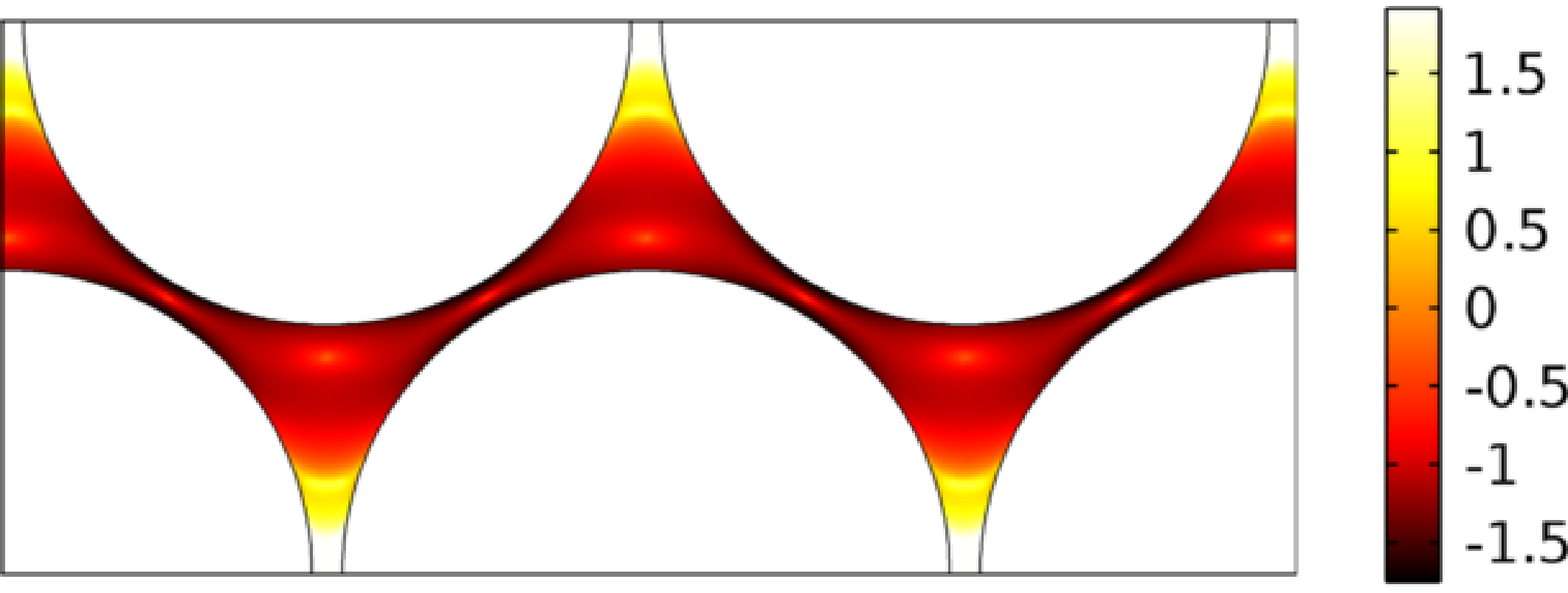}
    		    		    }
		      \end{subfigure}
		    \end{minipage}
		    \begin{subfigure}[]{
		    \includegraphics[width=0.45\textwidth]{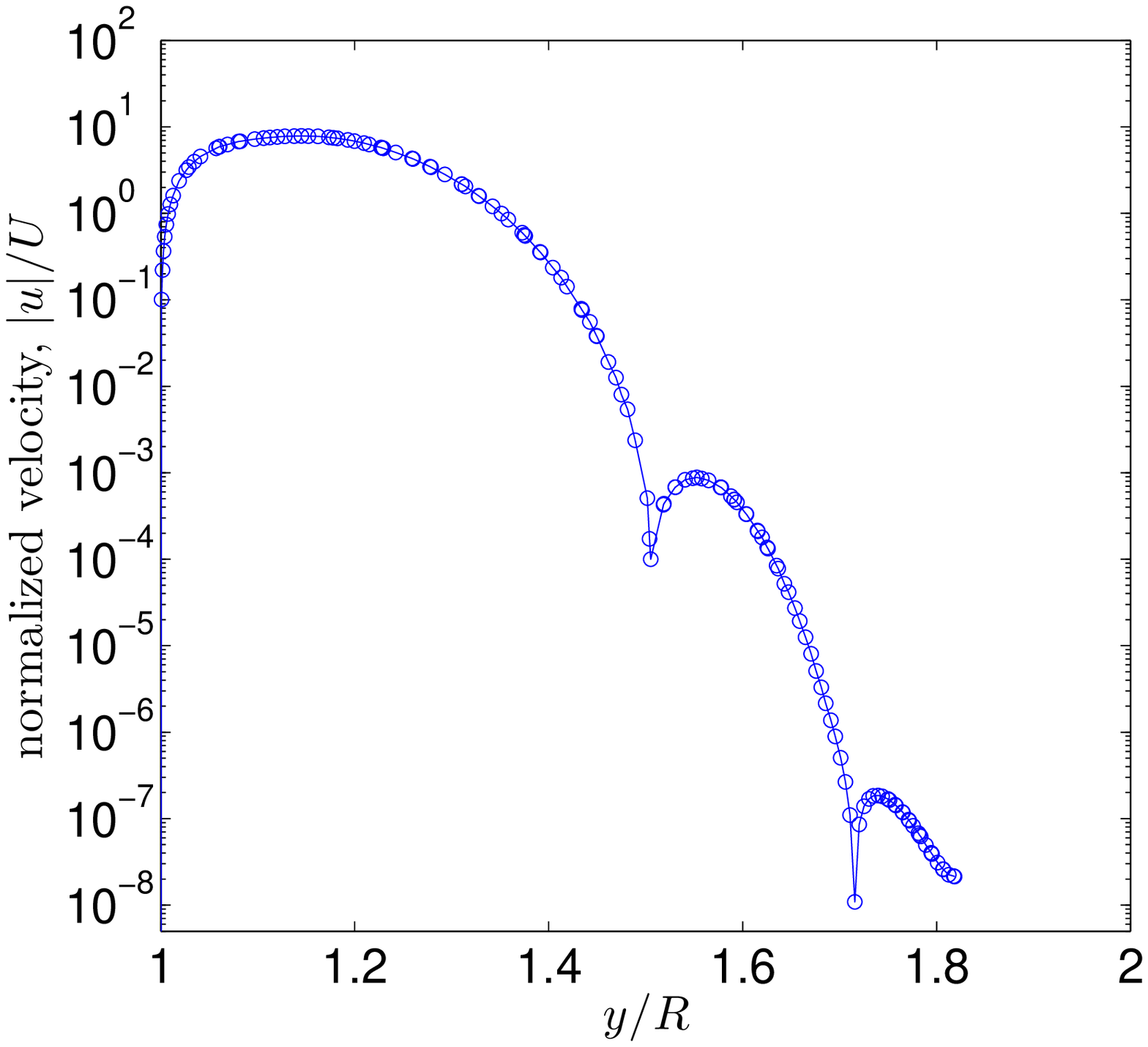}
  		    }
  		    \end{subfigure}
		    \hfill
		    \caption{Results from the numerical model for steady viscous flow of a shear-thinning fluid through fibers with hexagonal arrangement with porosity $\varepsilon=0.18$ and a shear-thinning fluid (with power-law exponent $n=0.5$): (a) contours of the normalized velocity magnitude $\|\textbf{u}\|/U$, (b) logarithmic contours of the dimensionless viscosity $\log\left(\eta\, m^{-1}\left(U/d\right)^{1-n}\right)$, the viscosity varies by a factor of $10^3$ from the high shear regions in between cylinders to the stagnant regions in the wake of the cylinder, (c) velocity distribution in the gap between the fibers (along the line marked white in (a)). The flow is in the $x$ direction.   }\label{fig:velocity-viscosity-hex}
		  \end{figure}
\FloatBarrier
	\section{Analytical model}
	In this section, we develop a locally-fully\textendash developed unidirectional flow model, which can be viewed as a lubrication analysis of the problem in cylindrical coordinates. We consider the unit cell shown in Figure \ref{fig:domain-analytical}, which represents the symmetrical domain for flow over an array of fibers with diameter $d=2R$ and spacing $s$. In transverse flow though a fibrous medium with a square arrangement of cylindrical fibers, the pressure drops periodically over the length of this unit cell, $s$ (the fiber spacing). In \cite{Bruschke1993}, a lubrication analysis in Cartesian coordinates has been applied, which assumes that the dominant velocity component is always parallel to the horizontal axis. This is clearly inappropriate for arbitrary spacing $s$. Here, instead of Cartesian coordinates, we use cylindrical polar coordinates and assume that the dominant velocity component is in the tangential direction. Hence, we can more reasonably capture the curvature in the streamlines near the narrow gap between cylinders (cf. Figure \ref{fig:streamlines-square}), which provides the main contribution to the total pressure drop across a unit cell.\\
	In cylindrical polar coordinates, we assume that $v_r \ll v_\theta$, therefore, $\textbf{u}\simeq v_{\theta}(r)\,\hat{\textbf{e}}_{\theta}$. This approximation fails for $\theta \to 0$ and $\theta \to \pi$, where the velocity is directed in the radial direction. However, the velocity in these regions is much smaller and the contribution of these regions to the total pressure drop is insignificant compared to the narrow gap region around $\theta=\pi/2$. The shear rate also becomes much smaller as $\theta\to 0$ or $\theta \to \pi$ compared to that in the narrow gap between the cylinders ($\theta=\pi/2$) since a smaller velocity variation ($O(U)$ vs. $O\left(Us/(s-d)\right)$) occurs over a larger length scale ($O(s)$ vs. $O(s-d)$). Thus to leading order, the only non-zero component of the deformation rate tensor is $\dot{\gamma}_{r\theta}$ and the shear rate (defined in equation (\ref{eqn:shear rate definition})) is calculated as
		\begin{align}
			\dot{\gamma}=\left|\dot{\gamma}_{r\theta}\right|,
		\end{align}	
	which is a direct consequence of postulating a velocity of form $\textbf{u}= v_{\theta}(r)\,\hat{\textbf{e}}_{\theta}$. The tangential component of the Cauchy momentum equation for slow steady viscous flow is thus simplified to
		\begin{align}
			\frac{1}{r^2}\frac{\partial}{\partial r}\left(r^2\tau_{r\theta}\right)=\frac{1}{r}\frac{\partial p}{\partial \theta}
		\label{eqn:Cauchy-tangential}
		\end{align}
		\begin{figure} [b!]
			\centering
			\includegraphics[width=0.45\textwidth]{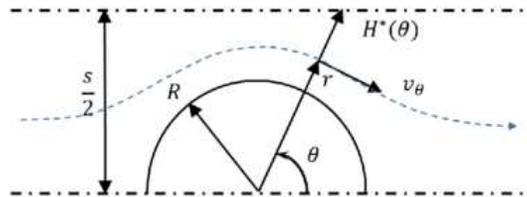}
			\caption[]{Schematic of the unit cell geometry used in the present analytical model, representing flow around a cylinder of diameter $d=2R$ and center-to-center cylinder spacing of $s$. The radial position of the symmetry line is denoted by $H^*(\theta)$.}
			\label{fig:domain-analytical}
		\end{figure}
	We assume that to leading order, the pressure $p$ is only a function of $\theta$ (i.e. $\frac{\partial p}{\partial \theta}=\frac{dp}{d\theta}$); hence, integration of equation (\ref{eqn:Cauchy-tangential}) yields the ${r\theta}$-component of the stress tensor:
		\begin{align}
			\tau_{r\theta}=\frac{c_0}{r^2}+\frac{1}{2} \frac{dp}{d \theta}
		\label{eqn:stress_r theta}
		\end{align}
	By substituting the constitutive equation for power-law fluids (equation (\ref{eqn:PL-viscosity})), into equation (\ref{eqn:stress_r theta}), we obtain a relation for the shear rate $\dot{\gamma}_{r\theta}$:
		\begin{align}
			\left|\dot{\gamma}_{r\theta}\right|^{n-1} \dot{\gamma}_{r\theta}=\frac{1}{2m} \frac{dp}{d \theta}\left(\frac{c}{{r^*}^2}-1\right)
		\label{eqn:shear_r theta_before_sign}
		\end{align}
	where $c$ is simply a rearranged form of the constant of integration, $c=-c_0\left(\frac{R^2}{2} \frac{dp}{d\theta}\right)^{-1}$, and $r^*=r/R$ is the dimensionless radial coordinate (with $R$ denoting the fiber radius: $R=d/2$).\\
	When evaluating the magnitude of the (rate-dependent) viscosity we deal with the absolute value of $\dot{\gamma}_{r\theta}$; so care must be taken in determining the sign of the terms before integration. With reference to Figure \ref{fig:domain-analytical}, if the bulk flow direction is from left to right ($v_{\theta} \le 0$), the condition on the pressure gradient should be $\frac{d p}{d \theta}\ge 0$. In this case, for $r^*\le \sqrt{c}$, we have $\dot{\gamma}_{r\theta}\ge 0$ and for $r^*\ge \sqrt{c}$, $\dot{\gamma}_{r\theta}\le 0$.\\
	Using the definition of $\dot{\gamma}_{r\theta}$, we obtain the following differential equation for the tangential velocity component,
		\begin{align}
			r^* \frac{\partial}{\partial r^*}\left(\frac{v^*}{r^*}\right) = \left\{ 
			  \begin{array}{l l}
			   \Pi^{1/n} \left(\frac{c}{r^{*^2}}-1\right)^{1/n} & \quad  \text{$\dot{\gamma}_{r\theta}\geq 0$ \,\,or\,\, $r^*\le\sqrt{c}$} \\
			    -\Pi^{1/n} \left(1-\frac{c}{r^{*^2}}\right)^{1/n} & \quad \text{$\dot{\gamma}_{r\theta}< 0$ \,\,or\,\, $r^*>\sqrt{c}$} \\
			  \end{array} \right.	
		\label{eqn:shear_r theta_after_sign}
		\end{align}
	where $\Pi$ is the dimensionless pressure gradient defined as 
		\begin{align}
			\Pi&\equiv\frac{R^n}{2mU^n}\frac{dp}{d\theta}.
			\label{eqn:Pi-definition}
		\end{align}
	The final velocity profile is derived by integrating equation (\ref{eqn:shear_r theta_after_sign}), which yields the following integral expression, in which the no slip condition $v^*=0$ on the cylinder at $r^*=1$ is also satisfied.
		\begin{align}
			v^*(r^*,\theta)&=\left\{ 
			\begin{array}{l l}
			\Pi^{1/n}r^*\int_{1}^{r^*}\frac{1}{\zeta}\left(\frac{c}{\zeta^2}-1\right)^{1/n}d\zeta & \quad  \text{$r^*\le\sqrt{c}$} \\
			 \Pi^{1/n}r^*\left[\int_{1}^{\sqrt{c}}\frac{1}{\zeta}\left(\frac{c}{\zeta^2}-1\right)^{1/n}d\zeta-\int_{\sqrt{c}}^{r^*}\frac{1}{\zeta}\left(1-\frac{c}{\zeta^2}\right)^{1/n}d\zeta\right] & \quad  \text{$r^*>\sqrt{c}$}  \\
				\end{array} \right.
			\label{eqn:velocity-integral form}
		\end{align}
	 The second boundary condition is symmetry at the centerline $r^*=H^*(\theta)$: $\partial v^*/\partial y\big|_{H^*(\theta)}=0$, where $H^*(\theta)$ is the dimensionless radial position of the symmetry line in cylindrical coordinates shown in Figure (\ref{fig:domain-analytical}) and calculated from equation (\ref{eqn:H definition}).  
		\begin{align}
			H^*(\theta)&=\frac{s}{d\sin\theta} \label{eqn:H definition}
		\end{align}
	The velocity gradient in the $y$-direction is obtained from projecting $\partial v^*/\partial r^*$ onto the $y$-axis; consequently, to satisfy the symmetry condition, $\partial v^*/\partial r^*\big|_{H^*(\theta)}=0$. Taking the derivative of the second expression in equation (\ref{eqn:velocity-integral form}) with respect to $r^*$ and setting it to zero at $r^*=H^*$ yields the following implicit equation for calculating $c$, the constant in equation (\ref{eqn:shear_r theta_before_sign}), as a function of $H^*$.
		\begin{align}
			\int_{1}^{\sqrt{c}}\frac{1}{\zeta}\left(\frac{c}{\zeta^2}-1\right)^{1/n} d\zeta-\int_{\sqrt{c}}^{H^*(\theta)}\frac{1}{\zeta}\left(1-\frac{c}{\zeta^2}\right)^{1/n} d\zeta-\left(1-\frac{c}{{H^*(\theta)}^2}\right)^{1/n}=0
		\label{eqn:c equation implicit}
		\end{align}
	The cell volumetric flow rate $Q_{\text{cell}}$  per unit depth of the fiber bed is calculated by integrating the tangential velocity profile from $r^*=1$ to $H^*(\theta)$. From mass conservation in the unit cell shown in Figure \ref{fig:domain-analytical}, we have
		\begin{align}
			Q_{\rm{cell}}=\frac{Ud}{2}\int_{1}^{H^*(\theta)}v^*dr^*=\frac{Us}{2},
			\label{eqn:Qcell}
		\end{align}
	which combined with equation (\ref{eqn:velocity-integral form}) leads to
		\begin{align}
			\Pi^{1/n}f(H^*,n)=\frac{s}{d} \label{eqn:Pi implicit}
		\end{align}
	where the dimensionless function $f(H^*,n)$ is defined as
		\begin{align}
			f(H^*,n)\equiv\Pi^{-1/n}\int_{1}^{H^*}v^*\, dr^*.
		\label{eqn:f(H)-definition}
		\end{align}
	Substituting equation (\ref{eqn:velocity-integral form}) into (\ref{eqn:f(H)-definition}) results in a double integral expression. Using integration by parts, we can convert the double integration to a single integration and write the following expression for $f(H^*,n)$.
		\begin{align}
			f(H^*,n)=\frac{1}{2}\int_{1}^{\sqrt{c}}\left(\frac{c}{\zeta^2}-1\right)^{1/n} \left(\frac{{H^*}^2}{\zeta}-\zeta\right)d\zeta - \frac{1}{2}\int_{\sqrt{c}}^{H^*}\left(1-\frac{c}{\zeta^2}\right)^{1/n} \left(\frac{{H^*}^2}{\zeta}-\zeta\right)d\zeta.
			\label{eqn:f(H)-general}
		\end{align}
	While this integral does not have an explicit solution for arbitrary real values of $n$, there exist explicit solutions for integer values of $\frac{1}{n}$, where $n$ is the power-law exponent for a shear-thinning fluid. Also for odd values of $\frac{1}{n}$ we can further simplify equation (\ref{eqn:f(H)-general}) by combining the two terms to a single term with integration limits from $1$ to $H^*$.\\ 
	Rearranging equation (\ref{eqn:Pi implicit}) yields the dimensionless pressure gradient (defined in equation (\ref{eqn:Pi-definition})) as:
		\begin{align}
			\Pi&\equiv\frac{R^n}{2mU^n}\frac{d p}{d \theta}=\left(\frac{s/d}{f(H^*,n)}\right)^n
			\label{eqn:Pi-result}
		\end{align}
	The pressure drop over one unit cell (with a length of $\Delta x=s$) is derived from using the definition of $\Pi$ from equation (\ref{eqn:Pi-definition}) and then integrating both sides of equation (\ref{eqn:Pi-result}) over a single cell:
		\begin{align}
			\Delta p_{\text{cell}}=\frac{mU^n}{d^n}2^{n+1}\left(\frac{s}{d}\right)^n\int_0^\pi \left(f\left(\frac{s}{d\sin\theta},n\right)\right)^{-n} d\theta
		\end{align}
	where $f$ is given in equation (\ref{eqn:f(H)-general}).\\
	Finally we obtain the dimensionless mobility defined in equation (\ref{eqn:mobility-nd-dfn}) from the expression
		\begin{align}
			M^*=\frac{\left(s/d\right)^{1-n}}{2^{n+1}\int_0^\pi \left(f(\frac{s}{d\sin\theta})\right)^{-n} d\theta}
			\label{eqn:mobility-analytical}
		\end{align}
	The simplest case is a Newtonian fluid, $n=1$, for which the function $f(H^*,n)$ has the following form:
		\begin{align}
			f(H^*,1)&=\frac{1}{4}\left({H^*}^2-2{H^*}^2\ln H^*-1\right) +\frac{1}{2} \frac{\left(\ln H^*+1\right)}{1+{H^*}^{-2}}\left({H^*}^2-2\ln H^*-1\right)
			\label{eqn:f(H)-Newtonian}
		\end{align}
	and the constant $c$ in the velocity profile is then calculated to be
		\begin{align}
			c|_{n=1}&=\frac{2{H^*}^2\left(\ln H^*+1\right)}{{H^*}^2+1}.  
		\end{align}
	For Newtonian fluids the dimensionless mobility is equivalent to the dimensionless permeability. Therefore, we obtain an analytical solution for permeability of fibrous media by setting ($n=1$) in equation (\ref{eqn:mobility-analytical}):
		\begin{align}
			M^*_{\rm{Newtonian}}=\kappa^*=\frac{1}{8}\left(\int_0^{\frac{\pi}{2}}\frac{1}{f\left(\sqrt{\frac{a}{1-\varepsilon}}\frac{1}{\sin\theta}\right)}d\theta\right)^{-1}
			\label{eqn:permeability-analytical}
		\end{align}
	Here we have used the symmetry expected at $\theta=\pi/2$ for integration and also converted the fiber spacing ratio $s/d$ (in the denominator of equation (\ref{eqn:mobility-analytical})) to porosity using the relationship $\varepsilon=1-a\,d^2/s^2 $. For a square arrangement of cylinders, the geometric parameter is $a=\frac{\pi}{4}$, while for a hexagonal arrangement $a=\frac{\pi}{2\sqrt{3}}$.\\ 
	Table \ref{table:velocity for PL} presents expressions the velocity profile for Newtonian fluids and power-law fluids with $n=0.5$ and $n=0.33$. The constant $c$ can be derived from equation (\ref{eqn:c equation implicit}). Figure \ref{fig:mobility-analytical} shows the dimensionless mobility at different porosities calculated from our numerical model in comparison with the derived analytical solution, equation (\ref{eqn:mobility-analytical}). We have focused here on power-law exponents of $n=1$, $\frac{1}{2}$ and $\frac{1}{3}$ in order to be able to calculate the velocity integrals in equation (\ref{eqn:f(H)-general}) explicitly. The method is, however, general and by developing an iterative procedure for finding the parameter $c$ for each angle $\theta$, we can solve the implicit equation (\ref{eqn:c equation implicit}) for arbitrary $n$ and then numerically evaluate the integrals in equations (\ref{eqn:f(H)-general}) and (\ref{eqn:mobility-analytical}). Figure \ref{fig:mobility-analytical} shows that shear-thinning can dramatically increase the dimensionless mobility of the fluid through the medium (especially for low porosities). It can be seen from Figure \ref{fig:mobility-analytical} that the analytical solution is in good agreement with our numerical data. For a shear-thinning fluid with power-law index $n=0.5$, at porosities $\varepsilon=0.35$ and $0.85$, the relative difference between the numerical data and analytical solution is $7 \%$ and $17 \%$ respectively, which is lower than the corresponding values for the traditional lubrication model ($18\%$ and $40\%$, respectively)\textemdash see \cite{Bruschke1993} and equation (\ref{eqn:mobility-analytical}).
	\begin{table}[h!]
	\center
	\caption{Analytical solution for the tangential velocity distribution in a unit cell, representing the space between fibers, for a Newtonian fluid $n=1$ and shear-thinning fluids with power-law exponents of $0.5$ and $0.33$. The constant $c$ is determined from equation (\ref{eqn:c equation implicit}).}
		\begin{tabular}{l*{3}{c}r}
			\hline
			$n$              & $v^*$ \\
			\hline
			$1$ 			 &  $-\Pi r^*\left(\frac{c}{2}-\frac{1}{2} \frac{c}{r^{*^2}}-\ln r^* \right)$  \\			
			$\frac{1}{2}$    & $-\Pi r^*\left(\frac{c^2}{4}-\frac{1}{4} \frac{c^2}{r^{*^4}}+\ln r^* +\frac{c}{r^{*^2}}-c\right)$ \\			
			$\frac{1}{3}$    & $-\Pi r^*\left(\frac{3}{4} \frac{c^2}{r^{*^4}}-\ln r^* -\frac{3}{2}\frac{c}{r^{*^2}}-\frac{1}{6}\frac{c^3}{r^{*^6}}-\frac{3}{4} c^2+\frac{3c}{2} +\frac{c^3}{6}\right)$ \\
			\hline
		\end{tabular}
	\label{table:velocity for PL}
	\end{table}
		\begin{figure}[h!]
			 \centering
			 \includegraphics[width=0.5\textwidth]{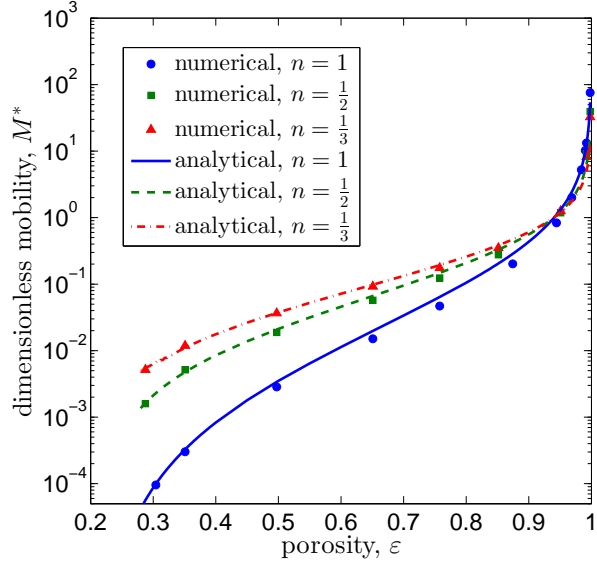}
			 \caption[]
			 {Comparison of dimensionless mobility $M^*$ for a square arrangement of fibers from numerical computation and analytical solution (equation (\ref{eqn:mobility-analytical})) at power-law exponents of $n=1$ , $n=1/2$, and $n=1/3$. The solid line ($n=1$) represents the analytical relationship for the dimensionless transverse permeability of an array of fibers, equation (\ref{eqn:permeability-analytical}).}\label{fig:mobility-analytical}
		\end{figure}
	\FloatBarrier
	\section{Scaling analysis} \label{sec:scaling analysis}
		Although our lubrication solution results in an analytic expression for the mobility, evaluation of this expression for arbitrary power-law exponent $n$ requires implicit solution of equation (\ref{eqn:c equation implicit}) and then numerical evaluation of two integrals. We show in this section that it is also possible to use a simple scaling analysis to obtain a relationship for the effective shear rate that characterizes the flow of an inelastic fluid with a rate-dependent viscosity through a fiber bed as a function of porosity. Since the main contribution to the total pressure drop is from the hydraulic resistance of the local constrictions in the flow domain, we estimate the shear rate by the following velocity gradient scale    
			\begin{align}
				\dot{\gamma}&\approx \frac{U_p}{\delta}
			\end{align}
		where $U_p$ is the characteristic velocity in the narrow gap between the cylinders and $\delta$ is the length scale in the gap over which the velocity varies. Using $\delta=(s-d)/2$ and $U_p=Us/(s-d)$, we get
			\begin{align}
				\dot{\gamma}_{\text{eff}}&=\frac{2U}{d} \frac{s/d}{(s/d-1)^2}
				\label{eqn:gamma-eff-sd}
			\end{align}
		If we non-dimensionalize the shear rate using $U/d$ to obtain $\dot{\gamma^*}=\dot{\gamma}_{\text{eff}} \left(\frac{U}{d}\right)^{-1}$, then the effective dimensionless shear rate in a fibrous medium can be expressed as a function of the porosity by substituting the relative fiber spacing $s/d$ in terms of the porosity using $\varepsilon=1-a d^2/s^2$:
			\begin{align}
				\dot{\gamma^*}=\frac{2\sqrt{a(1-\varepsilon)}}{\left(\sqrt{a}-\sqrt{1-\varepsilon}\right)^2} \label{eqn:gamma*}
			\end{align}
		where as before, $a=\pi/4$ for a square packing of fibers and $a=\pi/(2\sqrt{3})$ for a hexagonal packing.
		Subsequently, the effective value of the rate-dependent viscosity in the gap is
			\begin{align}
				\eta_{\rm{eff}}=\left(\frac{mU^{n-1}}{d^{n-1}}\right)\dot{\gamma^*}^{n-1}\label{eqn:eta_eff}
			\end{align}		
		and the dimensionless mobility is calculated as a function of the porosity and power-law exponent:
			\begin{align}
				M^*_{\text{scale}}=\kappa^*\left(\frac{2\sqrt{a(1-\varepsilon)}}{\left(\sqrt{a}-\sqrt{1-\varepsilon}\right)^2}\right)^{1-n}
				\label{eqn:mobility_nd_scale}
			\end{align}
		Here the dimensionless permeability, $\kappa^*$, is either known from experiments (with Newtonian fluids) or can be evaluated from either of the two approaches (numerical and analytical) developed in this study or from equation(\ref{eqn:permeability-tamayol}) of Tamayol and Bahrami \cite{Tamayol2011}:
			\begin{align}
				\kappa^*_{\text{theory}}\equiv\frac{\kappa}{d^2}=0.16\,a\frac{\left(1-a^{-1}\sqrt{1-\varepsilon}\right)^3}{\left(1-\varepsilon\right)\sqrt{\varepsilon}} \label{eqn:permeability-tamayol}
			\end{align}
			\begin{figure}[h!]
				 \centering
					\includegraphics[width=0.5\textwidth]{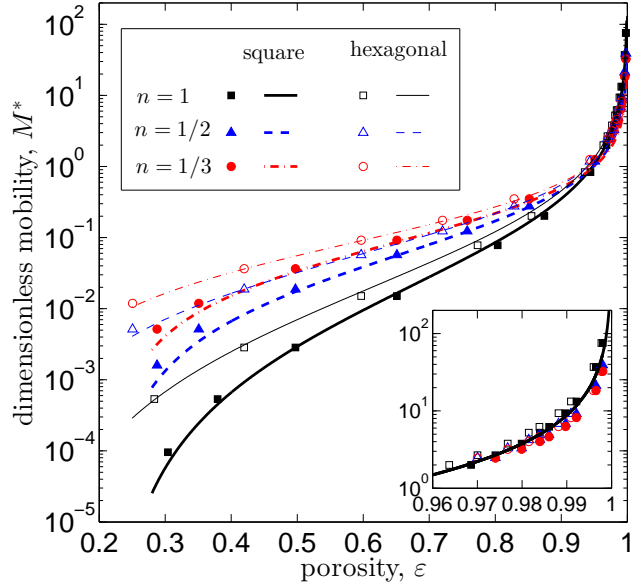}
				  \caption[]
				  {Dimensionless mobility as a function of porosity at 	different values of the power-law index for square and hexagonal fiber arrangements. The symbols are obtained from detailed numerical simulation and the lines represent the scaling result (equation (\ref{eqn:mobility_nd_scale}) and (\ref{eqn:permeability-tamayol})). The inset plot shows an expanded view of the main plot in the high porosity limit $\varepsilon\to 1$.}\label{fig:mobility-hex-square}
			\end{figure}
		This relationship is also derived from a scaling analysis and by balancing the local pressure drop with the shear stress at the pore scale. The coefficient 0.16 in Tamayol and Bahrami's relation was obtained from matching the permeability function with experimental and numerical data in the literature \cite{Tamayol2011}. \\
		Figure \ref{fig:mobility-hex-square} compares the results of the present scaling approach for flow transverse to an array of cylinders having either square or hexagonal arrangement with corresponding values obtained from numerical simulations at power-law exponents of $n=1$, $\frac{1}{2}$, and $\frac{1}{3}$. The two solid lines in this figure correspond to Newtonian fluids, for which the dimensionless mobility is equal to the dimensionless permeability (equation (\ref{eqn:permeability-tamayol}) \cite{Tamayol2011}), while the lines corresponding to $n=\frac{1}{2}$ and $\frac{1}{3}$ are the mobility for a power-law fluid evaluated from equation (\ref{eqn:mobility_nd_scale}) using the permeability $\kappa^*$ given by equation (\ref{eqn:permeability-tamayol}). Figure \ref{fig:mobility-hex-square} shows that this scaling approach yields good agreement with numerical simulations and can be applied to estimate the pressure\textendash velocity relation for power-law fluids in fibrous media. \\
		The impact of the specific fiber arrangement is most significant at low porosities but vanishes in the limit of very high porosities ($\varepsilon\to1$). This serves to re-emphasize that in modeling the mobility and permeability in fibrous media, effects of the tortuosity can not be neglected at medium and low porosities. According to the inset plot in figure \ref{fig:mobility-hex-square}, which shows an expanded view of the variation in the mobility with porosity, in the limit of high porosity ($\varepsilon\to1$), the dimensionless mobilities of all power-law fluids converge to the dimensionless permeability of the porous medium. Therefore, at very high porosities ($\varepsilon\gtrsim 0.95$), we can use the following simple equation to estimate the pressure drop due to the flow of a power-law fluid in a fibrous medium with permeability $\kappa$ and fiber diameter $d$:
			\begin{align}
				{\frac{\Delta p}{L}}\biggr\rvert_{\varepsilon\to 1} \approx\frac{mU^nd^{1-n}}{\kappa}
			\end{align}	
		where $\kappa$ is determined from equation (\ref{eqn:permeability-tamayol}) for a Newtonian fluid.
		This simple scaling approach can easily be extended to other generalized Newtonian fluids, such as Carreau fluids, to predict the pressure drop for flow through fibrous media. We present the methodology and results for Carreau fluids in the next section. 
	\section{Carreau fluids} \label{sec:Carreau fluids}
	The simple power-law model (equation (\ref{eqn:PL-viscosity})) has a well-known singularity as $\dot{\gamma}\to 0$ and we expect regions of vanishingly low shear rate along the symmetry lines of the geometry (perpendicular to the flow direction). The Carreau-Yasuda equation is a model that enables the description of the plateaus in viscosity that are observed experimentally when the shear rate approaches zero or infinity \cite{Bird1987}. The shear viscosity for this model is given by equation (\ref{eqn:Carreau-viscosity}).\\
	To investigate the effects of different rheological parameters on the pressure drop for steady flow of Carreau fluids transverse to a periodic array of cylinders, we used numerical simulations and varied control parameters such as the superficial velocity $U$, the material time constant $\lambda$, and the porosity of the geometry, keeping the pore-scale Reynolds number ($Re=\rho U d/\eta_{\text{eff}}$) below unity to eliminate effects of fluid inertia. Figure \ref{fig:pressure-Cu} shows the results of numerical computation for the dimensionless pressure drop as the Carreau number $Cu=\lambda U/d$ varies for different porosities. In these calculations, the cylinders have a square arrangement and the power-law exponent is $n=0.5$. Very similar trends are observed for different values of the exponent $n$ and geometric parameters.
		\begin{figure}[h!]
		  \centering
		  	\includegraphics[width=0.5\textwidth]{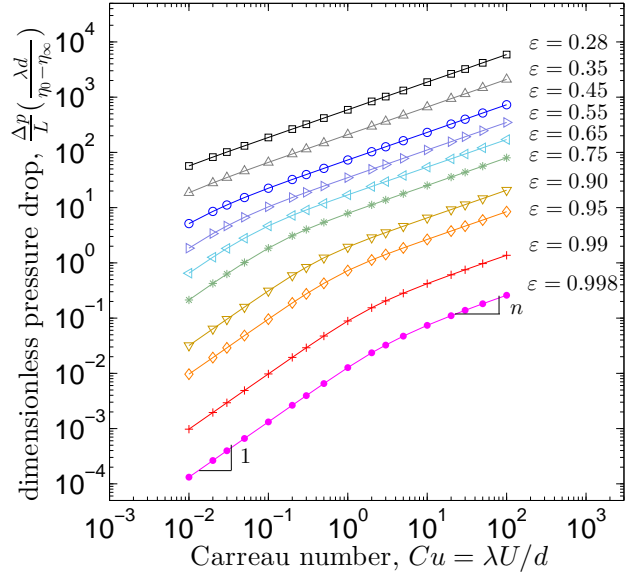}
		  \caption[]
		  {Effect of the Carreau number on the dimensionless pressure drop at different porosities from numerical model for a square arrangement of fibers with $n=0.5$. The lines are drawn to connect the numerical data points.}\label{fig:pressure-Cu}
		\end{figure}
	It can be observed in Figure \ref{fig:pressure-Cu} that at low porosities, the slope of the dimensionless pressure gradient $\left(\frac{\Delta p}{L}\frac{\lambda d}{\eta_0-\eta_\infty}\right)$ with the dimensionless velocity ($Cu$) is constant and equal to the exponent $n$, for a wide range of Carreau number. Since the local shear rate is relatively high in flows through low-porosity media, the fluid is always strongly sheared and the effective viscosity of the fluid is in the power-law region. However, as the porosity increases, the presence of a constant zero-shear-rate viscosity becomes important at lower Carreau numbers and for $Cu\ll1$ the material behaves like a Newtonian fluid with a pressure drop directly proportional to the velocity. At higher Carreau numbers ($Cu>1$), however, the fluid again transitions to a dominant power-law behavior.\\
	The self-similarity observed in the numerical data shown in Figure (\ref{fig:pressure-Cu}) suggests that for the flow of Carreau fluids through porous media, we should be able to identify an effective shear rate (which varies with $Cu$ and $\varepsilon$) that gives the effective viscosity that can be used in Darcy's law. The same scaling discussed in the previous section can be used to evaluate this effective shear rate. Substituting $\dot{\gamma}_{\text{eff}}$ from equation (\ref{eqn:gamma-eff-sd}) or (\ref{eqn:gamma*}) in equation (\ref{eqn:Carreau-viscosity}) yields the effective viscosity $\eta_{\text{eff}}$ for Carreau fluids, which can be used in the generalized Darcy law (equation (\ref{eqn:Darcy-law})):
		\begin{align}
			\frac{\Delta p}{L} =\frac{U\eta_{\text{eff}}}{\kappa}=\frac{U}{\kappa}\left(\eta_\infty+\left(\eta_0-\eta_\infty\right)\left(1+\left(\lambda \dot{\gamma}_{\rm{eff}}\right)^2\right)^\frac{n-1}{2}\right)
			\label{eqn:Carreau-pressure-velocity}
		\end{align}
	We use the Carreau number to non-dimensionalize the physical parameters in equation (\ref{eqn:Carreau-pressure-velocity}). Assuming that the infinite-shear-rate viscosity is negligible compared to the zero-shear rate viscosity $\eta_\infty\ll \eta_0$ (which holds for most shear-thinning materials), we can drop the first term in parentheses in equation (\ref{eqn:Carreau-pressure-velocity}). Rearranging this equation in a dimensionless form we obtain
		\begin{align}
			\frac{\Delta p}{L}\left(\frac{\lambda d}{\eta_0-\eta_\infty}\right)\kappa^*=Cu\left(1+\left(Cu \dot{\gamma}^*\right)^2\right)^{\frac{n-1}{2}}
			\label{eqn:Carreau-pressure}
		\end{align}
	The term $Cu\dot{\gamma}^*$  ($=\frac{\lambda U}{d} \dot{\gamma}^*$) in this equation motivates us to multiply both sides by $\dot{\gamma}^*$, so that the right hand side is only a function of a rescaled Carreau number $Cu\dot{\gamma}^*$, and define the following "scaled" dimensionless pressure drop:  
		\begin{align}
			\Pi_s=\frac{\Delta p}{L}\left(\frac{\lambda d}{\eta_0-\eta_\infty}\right)\kappa^* \dot{\gamma}^*=Cu \dot{\gamma}^*\left(1+\left(Cu \dot{\gamma}^*\right)^2\right)^{\frac{n-1}{2}}
			\label{eqn:Carreau-P-scaled}
		\end{align}
	This rescaled pressure drop enables superposition of the data shown in Figure \ref{fig:pressure-Cu} by scaling out the effect of porosity. Figure \ref{fig:pressure-Cu-collapsed} presents the scaled pressure drop, $\Pi_s$ versus the scaled Carreau number, $\frac{\lambda U}{d} \dot{\gamma}^*$. Note that we have used the permeability from the numerical model in calculating $\Pi_s$ in order to demonstrate the advantage of using the proposed scale for the dimensionless effective shear rate, $\dot{\gamma}^*$. The solid line in Figure \ref{fig:pressure-Cu-collapsed} is given by equation (\ref{eqn:Carreau-P-scaled}).
	 The dimensionless mobility of Carreau fluids predicted by the scaling model is calculated by rearranging equation (\ref{eqn:Carreau-P-scaled}) to yield:
		\begin{align}
			M^*=\kappa^*\left(Cu^{-2}+\dot{\gamma}^{*^2}\right)^{\frac{1-n}{2}}.
			\label{eqn:Carreau-mobility}
		\end{align}
	Note that $\kappa^*$ and $\dot{\gamma}^*$ are only functions of the relative fiber spacing (cf. equations (\ref{eqn:permeability-analytical}) and (\ref{eqn:gamma*}) respectively); therefore, the dimensionless mobility $M^*$ is a function of the Carreau number $Cu$, the power-law exponent $n$, and the relative fiber spacing $s/d$.	Figure \ref{fig:pressure-Cu-collapsed} shows that the proposed rescaling works extremely well in describing the flow of Carreau fluids in fibrous media and equation (\ref{eqn:Carreau-P-scaled}) successfully predicts the pressure drop\textendash flow rate relationship, especially when the permeability of the medium is known \text	it{a priori}.\\
	Figure \ref{fig:mobility-Cu} shows the variation of the dimensionless mobility with porosity at different Carreau numbers determined from the scaling model in comparison with the numerical simulation data computed for a Carreau fluid with a constant power-law exponent of $n=0.5$. In this figure we have evaluated the permeability $\kappa^*$ from equation (\ref{eqn:permeability-tamayol}) developed by Tamayol and Bahrami \cite{Tamayol2011}. Since this permeability is derived from a scaling analysis and fitting with numerical and experimental data at medium porosities, it does not perfectly capture the data in the very low porosity limit ($\varepsilon<0.35$). However, for higher porosities this expression is very accurate. Figure \ref{fig:mobility-Cu} also shows that for $\varepsilon\lesssim0.65$, all the values for the dimensionless mobility collapse on one curve and the fluid exhibits power-law behavior as a result of the high local shear rates in the pores. This asymptotic behavior for the Carreau model is once again described by equation (\ref{eqn:mobility_nd_scale}), in which the equivalent of the variable $m$ that appears in the dimensionless mobility for the power-law model (equation (\ref{eqn:mobility-nd-dfn})) is $m\equiv(\eta_0-\eta_{\infty})\lambda^{n-1}$. Deviation from this ubiquitous power-law behavior is only observed for $Cu<1$ and $\varepsilon\gtrsim0.65$. 
		\begin{figure}[h!]
			  \centering
			  \includegraphics[width=0.5\textwidth]{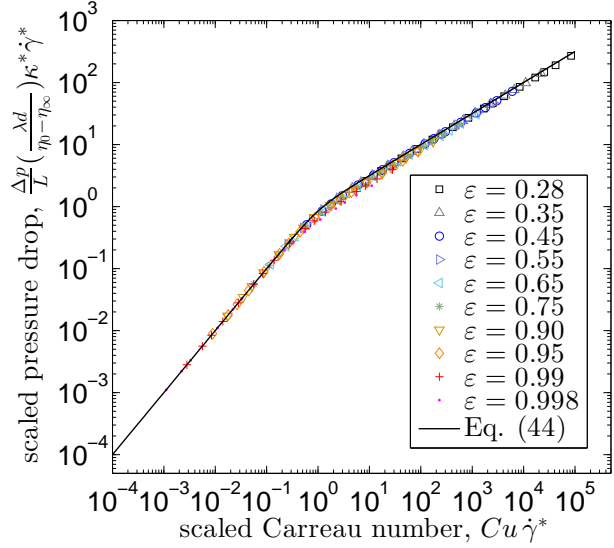}
			  \caption[]
			  {All of the numerical simulation data shown in Figure \ref{fig:pressure-Cu} can be superimposed to produce a master curve by an appropriate rescaling of  the Carreau number and the pressure drop. The line represents equation (\ref{eqn:Carreau-P-scaled}). }\label{fig:pressure-Cu-collapsed}
		\end{figure} 
		\begin{figure}[h!]
			  \centering
			  \includegraphics[width=0.5\textwidth]{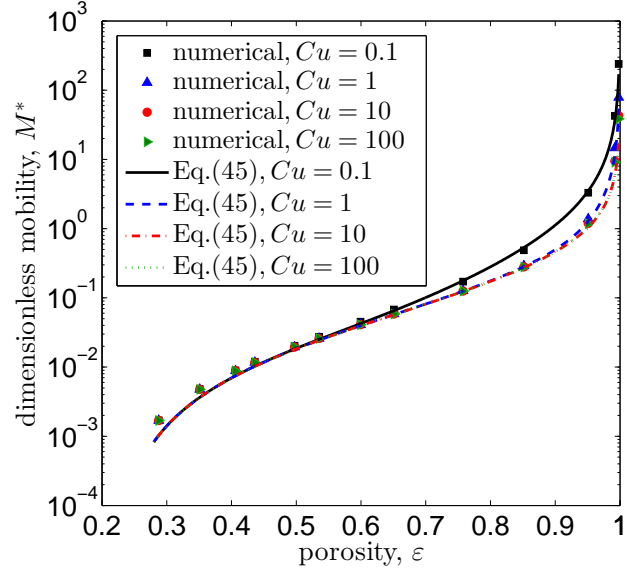}
			  \caption[]
			  {Dimensionless mobility of Carreau fluids as a function of porosity at different Carreau numbers for $n=0.5$. The symbols are numerical results and the lines are evaluated using equation (\ref{eqn:Carreau-mobility}) with permeability from equation (\ref{eqn:permeability-tamayol}). The universal asymptotic behavior observed at all values of $Cu$ for $\varepsilon<0.65$ is described by equation (\ref{eqn:mobility_nd_scale}). }. 
			  \label{fig:mobility-Cu}
		\end{figure}
\FloatBarrier	 
	\section{Conclusions}\label{sec:conclusions}	
	We have studied the steady viscous flow of shear-thinning power-law and Carreau fluids transverse to a periodic array of cylinders as a representative fibrous porous media. We have used three different approaches: numerical, analytical, and scaling analysis, to obtain the mobility function characterizing the flow of power-law fluids as a function of the flow rate, porosity of the fibrous medium, fiber diameter, and the rheological parameters that define the shear rate-dependent viscosity. This flow-rate-dependent mobility can be used to calculate the effective fluid viscosity in the system from equation (\ref{eqn:mobility-dfn}). \\
	For detailed pore-scale analysis of the flow we have used numerical simulations. Comparison of the results of the numerical simulations with the existing models in the literature (which are mainly developed for beds of particles) indicated the need for developing an improved theoretical model specific to flows of shear-thinning fluids through fibrous media. Consequently we developed analytical expressions for the transverse mobility function by assuming locally fully-developed flow of power-law fluids transverse to a periodic array of cylinders with a square arrangement. This analytical solution (equation (\ref{eqn:mobility-analytical})) agrees well with the numerical data for a wide range of porosity and other system parameters such as the power-law exponent $n$. To develop a more general theoretical model that can be applied to other fiber arrangements, we proposed a simple scaling analysis to evaluate the effective shear rate and viscosity characterizing the flow through the fibers. This simple scaling (equation (\ref{eqn:mobility_nd_scale})) can be used to predict the pressure drop as a function of the flow rate when the permeability of the medium is known (e.g. through experiments with a Newtonian fluid or from the theoretical results presented in this study).\\
	We extended our numerical studies and scaling analysis to also consider Carreau fluids, which allows us to capture the effects of a bounded zero-shear-rate viscosity. Based on parametric studies for a wide range of Carreau fluid rheology, we derived an appropriate criterion for transition from a constant viscosity fluid to a power-law fluid, which can be described by a rescaled Carreau number, $Cu^*=Cu \dot{\gamma}^*$ (with $\dot{\gamma}^*$ given by equation (\ref{eqn:gamma*}) as a function of porosity). We have shown that the power-law regime dominates when the rescaled Carreau number $Cu^*$ is greater than unity. In terms of the nominal or superficial Carreau number, we see that the fluid exhibits power-law behavior for all porosities when $Cu\gtrsim 1$ while for $Cu<1$, depending on the porosity of the medium, the fluid can be dominated by the constant-viscosity regime or the power-law regime. However, for $\varepsilon \lesssim 0.65$, all of the numerical results obtained from our Carreau simulations collapse onto a single curve given by our analysis for a simple power-law fluid (equation (\ref{eqn:mobility_nd_scale})) a result of the high local shear rates in the pores. A direct correspondence can be seen between this result and the recent experimental finding by Chevalier et al. \cite{chevalier2014breaking}, where it is shown that in the flow of yield stress fluids through porous media, the yielded regime is dominant for a medium with low or intermediate porosity due to the high local shear rates throughout the flow domain. The effective viscosity for $\varepsilon \lesssim 0.65$ can be simply calculated by
		\begin{align*}
			\eta_{\text{eff}}=m\left(\frac{U}{d}\right)^{n-1}\dot{\gamma}^{*^{n-1}}
		\end{align*}
	where the power-law viscosity coefficient $m$ for a Carreau fluid is $m=(\eta_0-\eta_{\infty})\lambda^{n-1}$. \\
	By comparing the three approaches, we suggest that for porosities $\varepsilon<0.9$, the more general form of our scaling model given by equation (\ref{eqn:Carreau-mobility}) can be used to predict the mobility of inelastic rate-dependent viscous fluids. This model is simple, yields reasonably good agreement with the numerical data, and can easily be extended to other generalized Newtonian fluid models.\\
	\\
	\textbf
	{Acknowledgment}
	\newline
	This work was supported in part by the MRSEC Program of the National Science Foundation under award number DMR - 0819762. The authors would also like to acknowledge Dr. Will Hartt and P\&G CETL for supporting complex fluids research in the Hatsopoulos Microfluids Laboratory. S.S. thanks Prof. Robert E. Cohen (MIT), Prof. Michael F. Rubner (MIT), and Prof. Brian L. Wardle (MIT) for helpful discussions.

		\bibliographystyle{IEEEtran}		
		\bibliography{IEEEabrv,biblio}
\end{document}